\documentclass[a4paper,11pt]{article}

\pdfoutput=1 

\usepackage{bbm}
\usepackage{jheppub} 
\usepackage[T1]{fontenc}
\usepackage{simplewick} 
\usepackage{makecell}	
\usepackage{tabu}		
\usepackage[makeroom]{cancel} 
\usepackage[normalem]{ulem} 
\usepackage[utf8]{inputenc}
\usepackage{amsmath,amsfonts,amsthm,amssymb} 
\usepackage{subcaption} 

\newcommand{\p}{\partial}
\newcommand{\eq}{&\quad}

\newcommand{\qRq}{\quad\Rightarrow\quad}
\newcommand{\rig}{\right.}
\newcommand{\lef}{\left.}

\newcommand{\para}{\parallel}

\newcommand{\tr}{\text{tr}}

\newcommand{\eff}{\text{eff}}
\newcommand{\disc}{\text{disc}}

\newcommand{\mco}{{\mathcal{O}}}
\newcommand{\D}{\mathcal{D}}
\newcommand{\cG}{\mathcal{G}}

\newcommand{\R}{\mathbb{R}}
\newcommand{\Z}{\mathbb{Z}}
\newcommand{\1}{\mathbbm{1}}

\newcommand{\hD}{{\hat{\Delta}}}

\newcommand{\hm}{{\hat{m}}}
\newcommand{\hO}{{\hat{\mathcal{O}}}}
\newcommand{\hp}{{\hat{\phi}}}

\newcommand{\f}{{(f)}}

\newcommand{\be}{\beta}
\newcommand{\ch}{\chi}
\newcommand{\de}{\delta}
\newcommand{\e}{\epsilon}
\newcommand{\ph}{\phi}
\newcommand{\g}{\gamma}

\newcommand{\la}{\lambda}
\newcommand{\m}{\mu}

\newcommand{\oo}{\omega}

\newcommand{\s}{\sigma}

\newcommand{\x}{\xi}

\newcommand{\De}{\Delta}

\newcommand{\G}{\Gamma}
\newcommand{\La}{\Lambda}

\preprint{UUITP-06/23}

\title{\boldmath The CW mechanism in a semi-definite system}

\author{Alexander Söderberg Rousu}

\affiliation{Department of Physics and Astronomy,
	Uppsala University,\\
	Box 516,
	SE-751 20 Uppsala,
	Sweden}

\emailAdd{alexander.soderberg.rousu@gmail.com}

\makeatletter
\gdef\@fpheader{}
\makeatother

\abstract{
	We study the $\phi^6 - \hat{\phi}^4$ model with $O(N)$-symmetry near three dimensions. This model has a sextic bulk-interaction and a quartic boundary-interaction. The bulk two-point correlator is found upto two-loops by solving the equation of motion and applying the boundary conditions. Finally we apply the Coleman-Weinberg mechanism to this model, which allows us to flow along the renormalization group to a first-ordered phase transition. At one-loop order only the boundary receives a non-trivial effective potential, giving the scalar on the boundary a vacuum expecation value. However, due to the boundary operator product expansion, the bulk one-point function is non-zero as well. This leads to a spontaneous symmetry breaking of the original $O(N)$-symmetry.

	All of the results in this paper was first presented in my thesis \cite{SoderbergRousu:2023ucv}.
	
}

\begin{document} 
	
\newtheorem{defin}{Definition}
\newtheorem{thm}{Theorem}
\newtheorem{cor}{Corollary}
\newtheorem{pf}{Proof}
\newtheorem{nt}{Note}
\newtheorem{ex}{Example}
\newtheorem{ans}{Ansatz}
\newtheorem{que}{Question}
\newtheorem{ax}{Axiom}

\maketitle

\section{Introduction}


Boundaries in quantum field theories are codimension one defects, where there is a physical region on only one side of it. Our understanding of \textit{boundary quantum field theory} (BQFT) is important to understand certain materials, e.g. graphene \cite{Teber:2012de}. 
Experimental setups in condensed matter will also have boundaries, making our understanding of them crucial. Moreover, they are important for understanding \textit{phase transitions} (p.t.'s) of materials, which is related to the \textit{boundary condition} (b.c.) on the boundary (see \cite{Diehl:1996kd, Herzog:2017xha, McAvity:1995zd} and references therein). E.g. the ordinary p.t., where the scalar satisfy Dirichlet b.c.'s, or the special p.t. with Neumann b.c.'s.




In $d = 3 - \e$ we can consider the $\ph^6 - \hp^4$ model satisfying $O(N)$-symmetry \cite{eisenriegler1988surface}, which is governed by the action
\begin{equation} \label{phi^6 - phi^4}
\begin{aligned}
S = \int_{\R^d_+}d^dx \left( \frac{(\p_\m\ph^i)^2}{2} + \frac{g_0}{48}\left[(\ph^i)^2\right]^3 \right) + \int_{\R^{d - 1}}d^{d - 1}x_\para \frac{\la_0}{8}\left[(\hp^i)^2\right]^2 \ .
\end{aligned}
\end{equation}
Here $\R^d_+ = \{ x_\para \in \R^{d - 1} ,\, x_\perp > 0 \}$, summation over the group indices, $i\in\{1, ..., N\}$, is implicit and boundary-local operators are hatted. This is an interesting BQFT as it has an \textit{Renormalization group} (RG) \textit{fixed point} (f.p.) where both the bulk- and the boundary-interaction are non-zero.\footnote{See \cite{Prochazka:2020vog} for a study at the f.p. where only the boundary-interaction is present.} Due to the boundary-interaction, the theory at this f.p. describe a special p.t. with a modified Neumann b.c. \cite{Diehl:2020rfx}. See \cite{Herzog:2020lel} for a large $N$ analysis of this model.


Due to the conformal symmetry at the f.p.'s, the bulk-bulk correlator is given by an unknown function, $F(\x)$, given in terms of the dimensionless cross-ratio, $\x$
\begin{equation} \label{corr 2}
\begin{aligned}
\langle\phi^i(x)\phi^j(y)\rangle &= A_d\de^{ij}\frac{F(\x)}{|4\,x_\perp y_\perp|^{\De_\ph}} \ , \quad \x = \frac{s^2}{4\,x_\perp y_\perp} \ .
\end{aligned}
\end{equation}
Here $A_d$ is a normalization constant, and $x_\perp$, $y_\perp$ are the coordinates normal to the boundary. In Sec. \ref{Ch: EOM} we will apply the bulk \textit{equation of motion} (e.o.m.) in the $\ph^6 - \hp^4$ model \eqref{phi^6 - phi^4}  order by order in the couplings to find $F(\x)$ as a perturbative expansion in these couplings. At each order we will have undetermined constants, which are fixed by the b.c.'s. This idea is not new in itself, and has been applied prior to this paper in \cite{Giombi:2020rmc, Giombi:2021cnr, Bissi:2022bgu}. However, in neither of these works a coupling constant in both the bulk and on the boundary was considered.

We are unaware of any results on the bulk $\ph - \ph$ correlator in 
beyond the free theory. Using the e.o.m. we are able to find this correlator upto $\mco(\e)$ (first order in $g_0$ and second order in $\la_0$) without much effort. Furthermore, we will study the boundary-limit of this correlator using the methods from \cite{Prochazka:2019fah} to read off the anomalous dimension of $\hp$.

We will also comment on the operators being exchanged in the bulk and boundary bootstrap channels \cite{Liendo:2012hy}, and read off the non-trivial \textit{operator product expansion} (OPE) coefficients. From this analysis we observe some technical issues of the analytical bootstrap method in \cite{Bissi:2018mcq, Dey:2020jlc, SoderbergRousu:2023nvd} that arises in three dimensions .


In Sec. \ref{Sec: CW def} we will study the \textit{Coleman-Weinberg (CW) mechanism} in the presence of a defect. It has not been worked out in general before this paper (as far as we are aware). The CW mechanism is a well-established method in a \textit{homogeneous QFT} (without a defect) which allows us to flow along the RG to a \textit{first-order phase transitions} starting from a conformal second-order one \cite{PhysRevD.7.1888}. In the process we also find the $\be$-functions upto one-loop \cite{yamagishi1981coupling}. Prior to this paper, the CW mechanism has been applied to a BQFT in \cite{Prochazka:2020vog} wherein only boundary-interactions were considered. Here we consider in addition a bulk-interaction.

We will go through how path integration for a general defect works, and how a \textit{vacuum expectation value} (v.e.v.) on the defect will stretch out into the bulk using the \textit{defect operator product expansion} (DOE) (the OPE between a bulk-local operator and the defect itself). This will in turn induce a \textit{spontaneous symmetry breaking} (SSB) both in the bulk and on the defect.

We will then apply the CW mechanism to the $\ph^6 - \hp^4$ model, and flow along the RG away from the conformal f.p.'s. This gives us an effective potential on the boundary (only taking into account one-loop effects) for a first-order p.t., where a SSB of the $O(N)$-symmetry occurs. Applying the Higgs mechanism tells us that there exist massless Goldstone modes invariant under $O(N - 1)$-transformations, and a Higgs mode with both a bulk and a boundary mass.


\section{Correlators from the equation of motion} \label{Ch: EOM}

In this Section we will find the bulk correlator in the $\ph^6 - \hp^4$ model \eqref{phi^6 - phi^4} upto two-loops using the e.o.m. Furthermore, we will write out the CFT data that enters in its bootstrap equation, and comment on issues of the discontinuity method \cite{SoderbergRousu:2023nvd} in odd dimensions.

\subsection{Renormalization group flow}

\begin{figure} 
	\centering
	\includegraphics[width=0.5\textwidth]{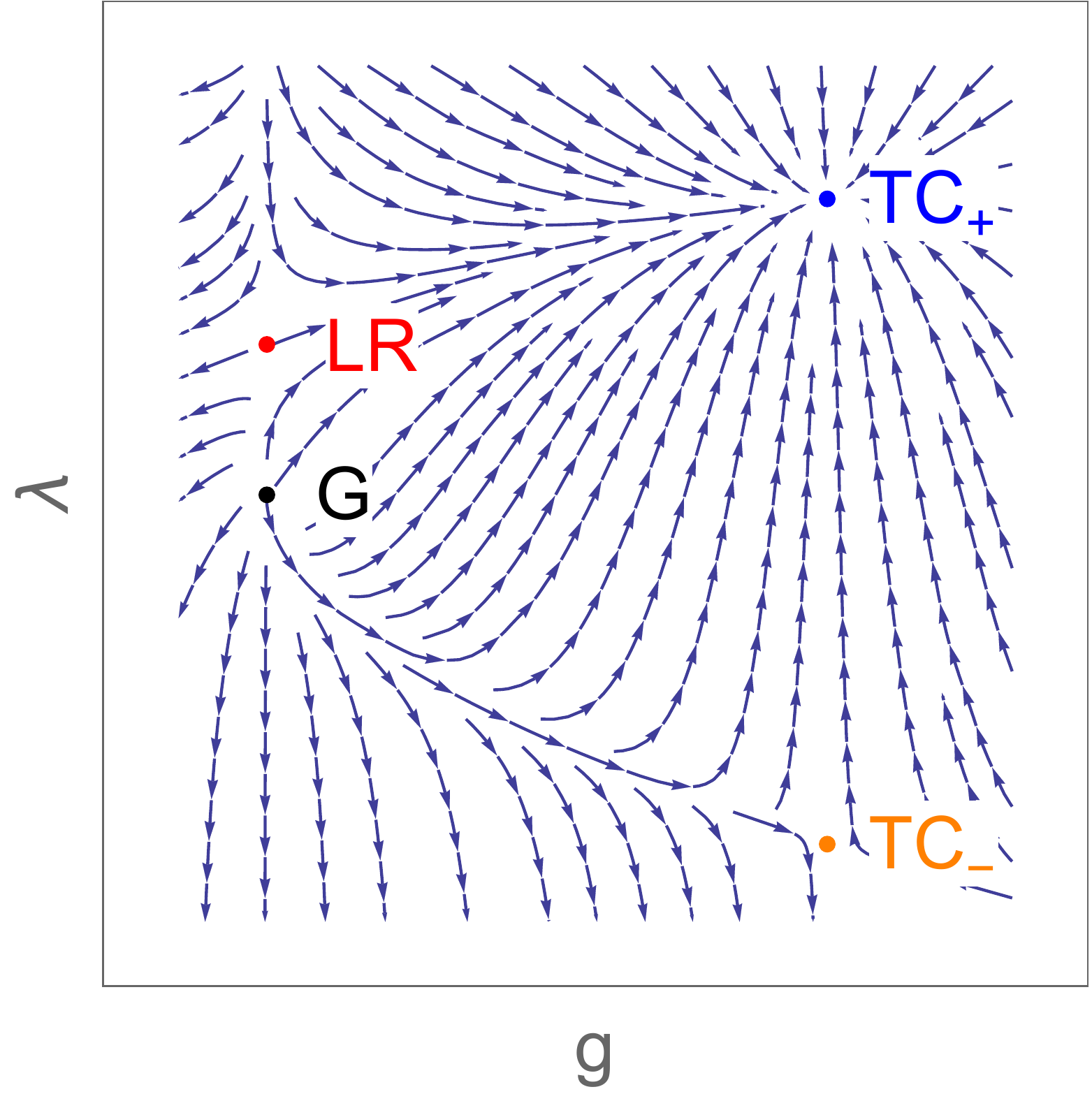}
	\caption{The RG flow of the $\ph^6 - \hp^4$ model \eqref{phi^6 - phi^4}. The black dot (G) is the trivial Gaussian f.p. \eqref{Gaussian fp}, the red one is the LR f.p. \eqref{LR fp}, the orange ($TC_-$) is the tricritical point \eqref{tri fp} with $\la^*_-$, and the blue ($TC_+$) that with $\la^*_+$. The blue f.p. is fully attractive.}
	\label{Fig: RGphi6phi4}
\end{figure}

The $\ph^6 - \hp^4$ model \eqref{phi^6 - phi^4} has the e.o.m. and b.c. (found by varying $\ph$ and $\hp$)
\begin{equation} \label{eom and bc}
\begin{aligned}
\p_\m^2\ph^i &= \frac{g_0}{8}\ph^4\ph^i \ , \quad \p_\perp\hp^i = \frac{\la_0}{2}\hp^2\hp^i \ ,
\end{aligned}
\end{equation}
where we have suppressed the summations over the $O(N)$-indices.\footnote{The boundary-local operator $\p_\perp\hp^i$ can be found from the boundary-limit of $\p_\perp\ph^i$ (in Sec. \ref{Sec: corr e} we will study this limit in more detail).} The $\be$-function for the bulk coupling is not affected by the boundary\footnote{Since bulk $\be$-functions only measure ultraviolet divergences in the coincident-limits of bulk-local fields, and not divergences in the near boundary-limit.} and thus it can be directly borrowed from the bulk $\ph^6$-theory \cite{duplantier1982lagrangian}
\begin{equation}
\begin{aligned}
\be_g &= -2\,\e\, g + \frac{3\,N + 22}{8\,\pi^2}g + \mco(g^2) \ .
\end{aligned}
\end{equation}
The boundary $\be$-function is (upto two-loops) \cite{eisenriegler1988surface, diehi1987walks}
\begin{equation} \label{eom bdy beta}
\begin{aligned}
\be_\la &= -\e\,\la - \frac{\pi(N + 4)}{8\,\pi}g + \frac{N + 8}{4\,\pi}\la^2 - \frac{(N + 4)(N - 62)}{64\,\pi^2}\la\,g  + \\
\eq - \frac{(5N + 22)\log 2}{2\,\pi^2}\la^3 + ... \ .
\end{aligned}
\end{equation}
Setting both of these $\be$-functions to zero gives us three RG f.p.'s. Aside from the trivial Gaussian f.p.
\begin{equation} \label{Gaussian fp}
\begin{aligned}
g_G^* \,,\ \la_G^*  = 0 \ , 
\end{aligned}
\end{equation}
there is also the \textit{long-range f.p.} (LR) studied in \cite{Prochazka:2019fah}
\begin{equation} \label{LR fp}
\begin{aligned}
g_{G}^* = 0 \ , \quad \la_{LR}^* = \frac{4\,\pi\,\e}{N + 8} + \frac{32\,\pi(5\,N + 22)\log\, 2}{(N + 8)^3}\e^2 + \mco(\e^3) \ ,
\end{aligned}
\end{equation}
and two f.p.'s where both the bulk and boundary couplings are non-trivial
\begin{equation} \label{tri fp}
\begin{aligned}
g^* &= \frac{16\,\pi^2\e}{3\,N + 22} + \mco(\e^2) \ ,
\end{aligned}
\end{equation}
\begin{equation*}
\begin{aligned}
\la_\pm^* &=  \pm 2\,\pi\sqrt{\frac{2(N + 4)\e}{(N + 8)(3\,N + 22)}} + \\
\eq + \frac{2\,\pi}{N + 8} \left( 1 + \frac{(N + 4)(N - 62)}{4(3\,N + 22)} + \frac{4(N + 4)(5\,N + 22)\log 2}{(N + 8)(3\,N + 22)} \right) \e + \mco(\e^{3/2}) \ .
\end{aligned}
\end{equation*}
We will mostly focus on the latter two f.p.'s. Here the bulk-interaction is at the \textit{tricritical point}, i.e. the point in the phase diagram (pressure vs. temperature) where three phases coexist (e.g. solid, liquid and gas). Note that at these tricritical points, $\la_\pm^*$ admits an expansion in $\sqrt{\e}$, while $g^*$ is expanded in $\e$. Moreover, there exist no f.p. where only the bulk-interaction is non-trivial. The RG flow is depicted in Fig. \ref{Fig: RGphi6phi4}, and the tricritical f.p. with $\la^*_+$ is the fully attractive one.

\subsection{The correlator upto $\mco(\sqrt{\e})$}

Let us consider the tricritical f.p. \eqref{tri fp} and find the bulk-bulk correlator. From conformal symmetry we know that it has to be on the form \eqref{corr 2}. Since $\langle\hp^2(x_\para)\rangle$ is trivial due to the conformal symmetry along the boundary, the e.o.m. and the b.c. \eqref{eom and bc} upto $\mco(\sqrt{\e})$ is
\begin{equation} \label{Free eom}
\begin{aligned}
\p_\m^2\langle\ph^i(x)\ph^j(y)\rangle &= \mco(\e) \ , \\
\langle\p_\perp\hp^i(x_\para)\ph^j(y)\rangle &= \frac{(N + 2)\la^*_\pm}{2}\langle\hp^2(x_\para)\rangle_\f\langle\hp^i(x_\para)\ph^j(y)\rangle_\f + \mco(\e) = \mco(\e) \ ,
\end{aligned}
\end{equation}
where the subscript $\f$ means the correlator from the free theory. At this order we are asked to solve the classical \textit{Klein-Gordon} (KG) eq. with Neumann b.c.. The bulk e.o.m. gives us
\begin{equation} \label{Diff}
\begin{aligned}
&\frac{A_d}{x_\perp^{\De_\ph + 2}(4\,y_\perp)^{\De_\ph}} \left( \x(\x + 1)F^{\prime\prime}(\x)  + \De_\ph(\De_\ph + 1)F(\x) + \rig \\
\eq \lef + \left( (\De_\ph + 1)(2\,\x + 1)  - \left(\De_\ph - \frac{d - 2}{2}\right) \frac{x_\perp}{y_\perp} \right) F^\prime(\x) \right) = 0 \ .
\end{aligned}
\end{equation}
Since $F(\x)$ should only depend on the cross-ratio, the $\frac{x_\perp}{y_\perp}$-term has to vanish. This puts a constraint on the scaling dimension of the bulk field
\begin{equation} \label{Free scaling dim}
\begin{aligned}
\De_\ph = \De_\ph^\f + \mco(\e) \ , \quad \De_\ph^\f = \frac{d - 2}{2} \ .
\end{aligned}
\end{equation}
This is indeed the correct result for a free bulk scalar, which should not get affected by the boundary coupling. The differential eq. \eqref{Diff} then yields
\begin{equation}
\begin{aligned}
4\,\x(\x + 1)F^{\prime\prime}(\x) + 2\,d(2\,\x + 1)F^\prime(\x) + d(d - 2)F(\x)  = 0 \ ,
\end{aligned}
\end{equation}
which can be solved upto two constants $A$ and $B$
\begin{equation}
\begin{aligned}
F(\x) = A\,\x^{-\De_\ph} + B(\x + 1)^{\De_\ph} \ .
\end{aligned}
\end{equation}
By applying the Neumann b.c. to \eqref{corr 2} we can fix one of the constants
\begin{equation}
\begin{aligned}
\frac{A_d\De_\ph y_\perp}{2(s_\para^2 + y_\perp^2)^{\frac{d}{2}}}(A - B) = 0 \qRq B = A \ .
\end{aligned}
\end{equation}
At this point we have
\begin{equation}
\begin{aligned}
F(\x) = A(\x^{-\De_\ph} + (\x + 1)^{\De_\ph}) \ .
\end{aligned}
\end{equation}
Since this has the same form of a free scalar, its conformal block decomposition is also the same. That is, in the bulk-channel only the identity operator and $\ph^2$ is exchanged, and in the boundary-channel only $\hp$ is exchanged \cite{Liendo:2012hy}
\begin{equation}
\begin{aligned}
\la^{\ph\ph}{}_{\1} &= A + \mco(\e) \ , \quad \la^{\ph\ph}{}_{\ph^2}\m^{\ph^2}{}_\1 = A + \mco(\e) \ , \quad (\m^\ph{}_\hp)^2 = 2\,A + \mco(\e) \ .
\end{aligned}
\end{equation}
Moreover, it tells us that neither $\ph^2$ nor $\hp$ receives an anomalous dimension 
\begin{equation}
\begin{aligned}
\g_{\ph^2} = \mco(\e) \ , \quad \g_{\hp} = \mco(\e) \ .
\end{aligned}
\end{equation}
The constant $A$ can be fixed by normalization, which we will chose to be
\begin{equation} \label{Normalization}
\begin{aligned}
\la^{\ph\ph}{}_{\1} = 1 \ , \quad \text{(exactly)} \qRq A = 1 \ ,
\end{aligned}
\end{equation}
which gives us
\begin{equation} \label{Free CFT data}
\begin{aligned}
F(\x) &= \x^{-\De_\ph} + (\x + 1)^{\De_\ph} \ , \quad \la^{\ph\ph}{}_{\ph^2}\m^{\ph^2}{}_\1 = 1 + \mco(\e) \ , \quad (\m^\ph{}_\hp)^2 = 2 + \mco(\e) \ .
\end{aligned}
\end{equation}
This is the same correlator as that found in a free theory, which is expected as it has no non-trivial Feynman diagrams at $\mco(\sqrt{\e})$.

\subsection{The correlator at $\mco(\e)$} \label{Sec: corr e}

At $\mco(\e)$, the e.o.m. \eqref{eom and bc} becomes
\begin{equation} \label{eom e}
\begin{aligned}
\p_\m^2\langle\ph^i(x)\ph^j(y)\rangle &= \frac{(N + 4)(N + 2)g^*}{8}\langle\ph^2(x)\rangle_\f^2\langle\ph^i(x)\ph^j(y)\rangle_\f + \mco(\e^{3/2}) \ , \\
\langle\p_\perp\hp^i(x_\para)\ph^j(y)\rangle &= \frac{\la^*_\pm}{2}\langle\hp^2\hp^i(x_\para)\ph^j(y)\rangle_{\la} + \mco(\e^{3/2}) \ .
\end{aligned}
\end{equation}
Here the subscript $\la$ denotes the correlator at $\mco(\la_\pm^*)$. In order to solve this differential equation we need $\langle\ph^2(x)\rangle_\f$ and
$\langle\hp^2\hp^i(x_\para)\ph^j(y)\rangle_{\la}$. 

Firstly let us find $\langle\ph^2(x)\rangle_\f$. It can be found from the $\ph \times \ph$ bulk OPE\footnote{Note that spinning operators have trivial one-point functions in the presence of a boundary \cite{Billo:2016cpy}.}
\begin{equation} \label{phi phi OPE}
\begin{aligned}
\langle\ph^i(x)\ph^j(y)\rangle_\f = A_d\de^{ij}\sum_{\mco}\frac{1}{|s|^{2\,\De_\ph}}C_{\De}(x)\langle\mco(x)\rangle_\f \ ,
\end{aligned}
\end{equation}
where we rescaled the exchanged bulk operator as $\mco \rightarrow \la^{\ph\ph}{}_\mco\, \mco$. We will compare this with the coincident-limit of the correlator (\ref{corr 2}, \ref{Free CFT data})
\begin{equation}
\begin{aligned}
\langle\ph^i(x)\ph^j(y)\rangle_\f = A_d\de^{ij} \left( \lim\limits_{y \rightarrow x} \frac{1}{s^{\De_\ph}} + \frac{1}{|2\,x_\perp|^{2\,\De_\ph}} \right) \ ,
\end{aligned}
\end{equation}
from which we can see that the first term corresponds to the identity exchange, and the second to the v.e.v. of $\ph^2$
\begin{equation}
\begin{aligned}
\langle\ph^2(x)\rangle_\f = A_d\frac{\m^{\ph^2}{}_{\1}}{|2\,x_\perp|^{2\,\De_\ph}} \ , \quad \m^{\ph^2}{}_{\1} = 1 \ .
\end{aligned}
\end{equation}
This is on the form we expect from conformal symmetry \cite{Billo:2016cpy}, and is consistent with the CFT data in \eqref{Free CFT data}.\footnote{We rescaled the one-point function with a factor $2^{-2\,\De_\ph}$ for simplicity.} 

\begin{figure} 
	\centering
	\includegraphics[width=0.5\textwidth]{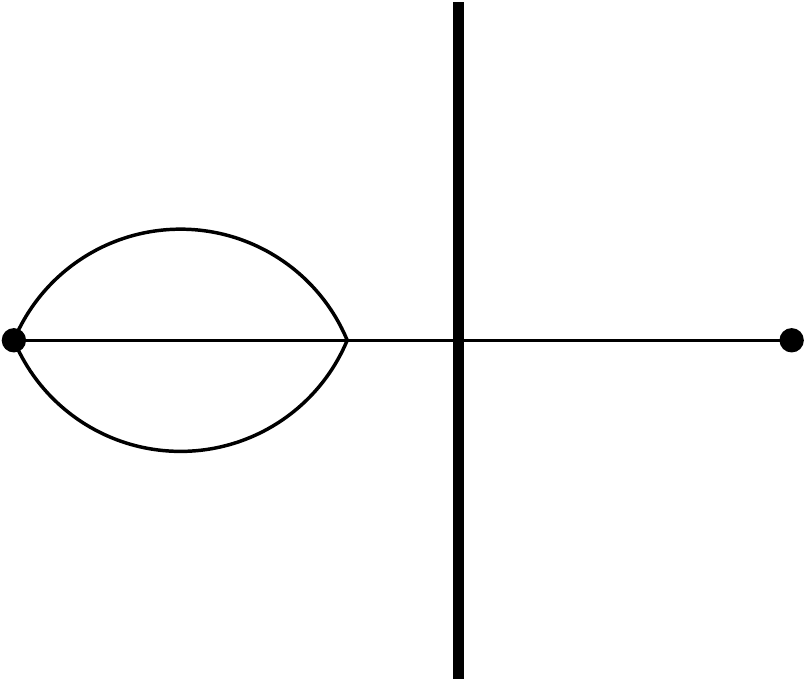}
	\caption{The diagram at $\mco(\la)$ for the $\hp^3-\ph$ correlator, where the LHS of the thick line represents the boundary, and the RHS the bulk.}
	\label{Fig: phi^3 - phi}
\end{figure}

Now we need to find 
$\langle\hp^2\hp^i(x_\para)\ph^j(y)\rangle_{\la}$. It is given by the Feynman diagram in Fig. \ref{Fig: phi^3 - phi}
\begin{equation*} 
\begin{aligned}
\langle\hp^2\hp^i(x_\para)\ph^j(y)\rangle_{\la} &= -16(N + 2)\la_\pm^*A_d\de^{ij}  \int_{\R^{d - 1}}d^{d - 1}z_\para \frac{1}{|x_\para - z_\para|^{3\,\De_\ph}[(z_\para - y_\para)^2 + y_\perp^2]^{\De_\ph} } \\
&= -16(N + 2)\la_\pm^*A_d\de^{ij} J^{d - 1}_{3\,\De_\ph, \De_\ph}(-s_\para, y_\perp^2) \ .
\end{aligned}
\end{equation*}
In the second line we performed the shift $z_\para \rightarrow z_\para + x_\para$, and wrote it in terms of the following master integral
\begin{equation} \label{Master int 2}
\begin{aligned}
J^n_{a,b}(z, w^2) &\equiv \int_{\R^n}\frac{d^nx}{x^{2\,a}[(x - z)^2 + w^2]^b} =  \\
&= \frac{\G_{a + b}}{\G_a\G_b}\int_{0}^{1}du(1 - u)^{a - 1}u^{b - 1}\int_{\R^n}\frac{d^nx}{(x^2 + u(1-u)z^2 + u\,w^2)^{a + b}} \\
&= \frac{\pi^{\frac{n}{2}}\G_{a + b - \frac{n}{2}}\G_{\frac{n}{2} - a}}{\G_b\G_{\frac{n}{2}}(z^2 + w^2)^{a + b - \frac{n}{2}}}{}_2F_1\left( a + b - \frac{n}{2}, n - \frac{a}{2}, \frac{n}{2}, \frac{z^2}{z^2 + w^2} \right)  \\
&= \frac{\pi^{\frac{n}{2}}\G_{a + b - \frac{n}{2}}\G_{\frac{n}{2} - a}\G_{\frac{n}{2} - b}}{\G_a\G_b\G_{n - a - b}}\frac{1}{|z|^{2(a + b) - n}} \ , \quad \text{if $w = 0$.}
\end{aligned}
\end{equation}
Here $\G_x = \G(x)$ is a shorthand notation for the Gamma function. In going from the second to the third row we used a Schwinger parametrization 
\begin{equation} \label{Sch param}
\begin{aligned}
\frac{1}{A^n} = \int_0^\infty du\frac{u^{n - 1}}{\G_n}e^{-u\,A} \ ,
\end{aligned}
\end{equation}
followed by a Gaussian integration in $\R^n$
\begin{equation}
\begin{aligned}
\int_{\R^n}d^nx e^{-a\,x^2 + b\,x} = \left( \frac{\pi}{a} \right)^{\frac{n}{2}}\exp\left(\frac{b^2}{4\,a}\right) \ .
\end{aligned}
\end{equation}
It gives us
\begin{equation} \label{bc rhs}
\begin{aligned}
\langle\hp^2\hp^i(x_\para)\ph^j(y)\rangle_{\la_\pm^*} &= A_d\de^{ij}\m^\ph{}_{\hp^3}\frac{y_\perp}{(s_\para^2 + y_\perp^2)^{\frac{3}{2}}} + \mco(\e) \ , \\
\m^\ph{}_{\hp^3} &= \frac{(N + 2)\la_\pm^*}{2\,\pi^2} + \mco(\e) \ .
\end{aligned}
\end{equation}
This has the expected form from conformal symmetry \cite{Billo:2016cpy}.

We have now everything we need to solve the e.o.m. \eqref{eom e} at $\mco(\e)$. We will consider the $\e$-expansion
\begin{equation} \label{e exp dyn}
\begin{aligned}
\De_\ph &= \frac{d - 2}{2} + \e\,\g_\ph + \mco(\e^2) \ , \quad F(\x) = F_{(0)}(\x) + \e\, F_{(1)}(\x) + \mco(\e^{\frac{3}{2}}) \ ,
\end{aligned}
\end{equation}
where $F_{(0)}(\x)$ is the solution \eqref{Free CFT data} upto $\mco(\sqrt{\e})$. By expanding the bulk e.o.m. in $\e$ we find a differential equation for $F_{(1)}(\x)$
\begin{equation} \label{Diff eq. e}
\begin{aligned}
0 &= \frac{\de^{ij}\e}{8\,\pi\, x_\perp^{\frac{5}{2}}\sqrt{y_\perp}} \left\{ \x(\x + 1) F_{(1)}^{\prime\prime}(\x) + \frac{3(2\,\x + 1)}{2} F_{(1)}^{\prime}(\x) + \frac{3}{4} F_{(1)}(\x) + \rig \\
\eq - \frac{1}{2} \left( \frac{(N + 2)(N + 4)g^*}{256\,\pi^2\e} - \g_\ph \right) \left( \frac{1}{\sqrt{\x}} + \frac{1}{\sqrt{\x + 1}} \right) + \\
\eq \lef - \g_\ph \left[ \left( \frac{1}{\x^{\frac{3}{2}}} - \frac{1}{(\x + 1)^{\frac{3}{2}}} \right) + \frac{x_\perp}{y_\perp} \left( \frac{1}{\x^{\frac{3}{2}}} + \frac{1}{(\x + 1)^{\frac{3}{2}}} \right) \right] \right\}  \ .
\end{aligned}
\end{equation}
If we demand that $F_{(1)}(\x)$ should only depend on the cross-ratio $\x$, then we need to set the $\frac{x_\perp}{y_\perp}$-term to zero. It yields
\begin{equation} \label{Anom dim phi}
\begin{aligned}
\g_\ph &= 0 \ .
\end{aligned}
\end{equation}
This agrees with the older literature \cite{duplantier1982lagrangian}. It can also be seen by assuming that $\De_\ph$ is not affected by the boundary coupling $\la$. Then in the case without a boundary there are no non-trivial Feynman diagrams for the $\ph - \ph$ correlator at $\mco(g)$. 

Above constraint brings \eqref{Diff eq. e} into
\begin{equation*}
\begin{aligned}
0 &= \x(\x + 1) F_{(1)}^{\prime\prime}(\x) + \frac{3(2\,\x + 1)}{2} F_{(1)}^{\prime}(\x) + \frac{3}{4} F_{(1)}(\x)  - \frac{(N + 2)(N + 4)g^*}{512\,\pi^2\e} \left( \frac{1}{\sqrt{\x}} + \frac{1}{\sqrt{\x + 1}} \right) \ ,
\end{aligned}
\end{equation*}
which has the solution
\begin{equation} \label{F1}
\begin{aligned}
F_{(1)}(\x) &= \frac{(N + 2)(N + 4)g^*}{256\,\pi^2\e} \left( \frac{1}{\sqrt{\x}} + \frac{1}{\sqrt{\x + 1}} \right) \log(\sqrt{\x} + \sqrt{\x + 1}) + \\
\eq + \frac{A}{\sqrt{\x}} + \frac{B}{\sqrt{\x + 1}} \ .
\end{aligned}
\end{equation}
There will be a $x_\perp$-pole in the boundary-limit of $\p_\perp\ph^i(x)$, and thus we cannot naively apply the b.c. \eqref{eom e}. To understand the origin of this pole we study the \textit{boundary operator product expansion} (BOE) of $\p_\perp\ph^i$ (which is a descendant of $\ph^i$ in the bulk). The BOE is the OPE between a bulk-local field and the boundary itself. For a scalar, it is given by \cite{Billo:2016cpy}
\begin{equation} \label{BOE}
\begin{aligned}
\mco(x) &= \sum_{\hO}\frac{\mu^{\mco}{}_{\hO}}{|x_\perp|^{\De - \hD}}\hat{C}(x_\perp^2\p_\para^2)\hO(x_\para) \ , \quad \hat{C}(x) = \sum_{m\geq 0}\frac{x^m}{(-4)^mm! \left( \hD - \frac{d - 3}{2} \right)_m} \ ,
\end{aligned}
\end{equation}
where the sum runs over boundary-local primaries, and the differential operator $\hat{C}$ generates the towers of descendants. The BOE of $\p_\perp\ph^i$ is thus
\begin{align} \label{phi phi BOE}
\langle\p_\perp\ph^i(x)\ph^j(y)\rangle &= A_d\de^{ij}\sum_{\hO}\p_{x_\perp}\frac{\m^\ph{}_\hO}{|x_\perp|^{\De_\ph - \hD}}\hat{C}(x_\perp^2\p_{x_\para}^2)\langle\hO^i(x_\para)\ph^j(y)\rangle \\
&= A_d\de^{ij}\sum_{\hO} \left( \frac{(\hD - \De_\ph)(\m^\ph{}_\hO)^2}{|x_\perp y_\perp|^{\De_\ph - \hD + 1}(s_\para^2 + y_\perp^2)^\hD} + \mco(x_\perp^{1 - \De_\ph + \hD}) \right) \ . \nonumber
\end{align}
Here we plugged in the form of the bulk-boundary correlator \cite{Billo:2016cpy} and expanded in $x_\perp$. To the lowest orders in $x_\perp$ the boundary fields $\hp$ and $\p_\perp\hp$ are exchanged. In the $\e$-expansion of their CFT data we have
\begin{equation}
\begin{aligned}
(\m^\ph{}_\hp)^2 &= 2 + \mco(\e) \ , \quad &\De_\hp &= \De_\ph^\f + \e\, \g_\hp + \mco(\e^\frac{3}{2}) \ , \\
(\m^\ph{}_{\p_\perp\hp})^2 &= \e\,\de\m + \mco(\e^\frac{3}{2}) \ , \quad &\De_{\p_\perp\hp} &= \De_\ph^\f + 1 + \mco(\e) \ ,
\end{aligned}
\end{equation}
where we remind the reader that $\De_\ph^\f$ is the free scaling dimension \eqref{Free scaling dim}. Plugging this into \eqref{phi phi BOE} yields
\begin{equation}
\begin{aligned}
\langle\p_\perp\ph^i(x)\ph^j(y)\rangle &= \frac{\e\, A_d\de^{ij}\g_\hp}{x_\perp\sqrt{s_\para^2 + y_\perp}} + \langle\p_\perp\hp^i(x_\para)\ph^j(y)\rangle + \mco(\e^\frac{3}{2}, x_\perp) \ , \\
\langle\p_\perp\hp^i(x_\para)\ph^j(y)\rangle &= \e\, A_d\de^{ij}\frac{y_\perp}{(s_\para^2 + y_\perp)^\frac{3}{2}}(\m^\ph{}_{\p_\perp\hp})_{\e}^2  + \mco(\e^\frac{3}{2}) \ .
\end{aligned}
\end{equation}
So from the pole in $x_\perp$ we can read off the anomalous dimension of $\hp$, while the $x_\perp^0$-term should be matched with the b.c. in \eqref{eom e}.

For the solution \eqref{F1} this yields
\begin{equation} \label{Bdy anom dim}
\begin{aligned}
\g_\hp &= -\frac{(N + 2)(N + 4)g^*}{512\,\pi^2\e} \ , \quad \de\m = A - B \ .
\end{aligned}
\end{equation}
The b.c. \eqref{eom e} with the BOE coefficient \eqref{bc rhs} now gives us
\begin{equation}
\begin{aligned}
B = A - \frac{(N + 2)(\la^*_\pm)^2}{4\,\pi^2\e} \ .
\end{aligned}
\end{equation}
Finally we normalize the theory according to \eqref{Normalization}
\begin{equation}
\begin{aligned}
A = 0 \ .
\end{aligned}
\end{equation}
This gives us
\begin{equation} \label{dyn e}
\begin{aligned}
F_{(1)}(\x) &= \frac{(N + 2)(N + 4)g^*}{256\,\pi^2\e} \left( \frac{1}{\sqrt{\x}} + \frac{1}{\sqrt{\x + 1}} \right) \log(\sqrt{\x} + \sqrt{\x + 1}) + \\
\eq - \frac{(N + 2)(\la^*_\pm)^2}{4\,\pi^2\e\sqrt{\x + 1}} \ .
\end{aligned}
\end{equation}
Note that the $\la^*_\pm$-term breaks the \textit{image symmetry} ($x_\perp \rightarrow -x_\perp$ or $\x \rightarrow \x + 1$). We retrieve the same $\ph - \ph$ correlator as in \cite{Prochazka:2019fah} when $g^* \rightarrow 0$, $\la^*_\pm \rightarrow \la^*_{LR}$ and $\e \rightarrow \e^2$. Note that by using the e.o.m. we just had to calculate one tree-level Feynman diagram for the $\hp^3 - \ph$ correlator at $\mco(\sqrt{\e})$ rather than two two-loop $\ph - \ph$ diagrams at $\mco(\e)$.

\subsection{Conformal block decomposition} \label{Sec: block decomp}

In this Section we will decompose the function \eqref{dyn e} in conformal blocks, and read off the OPE coefficients. At the RG f.p. we have an emerging conformal symmetry given by $SO(d, 1)$ (in the presence of a boundary in Euclidean space). This conformal symmetry tells us that the scalar correlator in the bulk satisfy the following bootstrap equation \cite{Liendo:2012hy}
\begin{equation}
\begin{aligned}
f(\x) = \x^{\De_\ph}F(\x) = \sum_{\mco}\la^{\ph\ph}{}_\mco \m^\mco{}_\1 \cG_\text{bulk}(\De; \x) = \x^{\frac{\De_{12}^+}{2}}\sum_{\hO}(\m^{\ph}{}_\hO)^2\cG_\text{bndy}(\hD; \x) \ ,
\end{aligned}
\end{equation}
wherein the \textit{bulk-channel} (with the conformal blocks $\cG_\text{bulk}$) scalar bulk-local operators, $\mco$, are exchanged, and in the \textit{boundary-channel} (with the conformal blocks $\cG_\text{bndy}$) scalar boundary-local operators, $\hO$, are exchanged. The bulk OPE coefficients, $\la^{\ph\ph}{}_\mco \m^\mco{}_\1$, are the bulk-bulk OPE coefficients, $\la^{\ph\ph}{}_\mco$, times the identity exchange, $\m^\mco{}_\1$, in the BOE \eqref{BOE}. In the boundary-channel, we have the square of the BOE coefficients, $(\m^{\ph}{}_\hO)^2$. The conformal blocks are known in closed form \cite{McAvity:1995zd}
\begin{equation}
\begin{aligned}
\cG_\text{bulk}(\De; \x) &= \x^{\De/2}{}_2F_1\left( \frac{\De}{2}, \frac{\De}{2}, \De - \frac{d - 2}{2}, -\x \right) \ , \\
\cG_\text{bndy}(\hD; \x) &= \x^{-\hD}{}_2F_1\left( \hD, \hD - \frac{d - 2}{2}, 2\,\hD - d + 2, -\x^{-1} \right) \ .
\end{aligned}
\end{equation}
For more details on conformal symmetry and bootstrap in the presence of a boundary, we direct the reader to \cite{Liendo:2012hy, Billo:2016cpy}. 

We consider the following $\e$-expansions
\begin{equation}
\begin{aligned}
\la^{\ph\ph}{}_{\ph^{2\,n}}\m^{\ph^{2\,n}}{}_\1 &\equiv \de_{n,1} + \e\,\de\la_n + \mco(\e^\frac{3}{2}) \ , \quad (\m^\ph{}_{\p_\perp^m\hp})^2 \equiv 2\,\de_{m,0} + \e\,\de\m_m + \mco(\e^\frac{3}{2}) \ .
\end{aligned}
\end{equation}
Starting with the bulk-channel, we note that the conformal blocks for $\ph^{2\,n}$, $n\in\{1, 2, 3\}$, are given by
\begin{equation}
\begin{aligned}
\cG_\text{bulk}(\De_{\ph^2}; \x) &= \sqrt{\frac{\x}{\x + 1}} - \frac{\e}{2} \left( \log\left( \frac{\x}{\x + 1} \right) - \frac{\g_{\ph^2}g^*}{\e}\log\x \right) + \mco(\e^{\frac{3}{2}}) \ , \\
\cG_\text{bulk}(\De_{\ph^4}; \x) &= \sqrt{\frac{\x}{\x + 1}}\log(\sqrt{\x} + \sqrt{\x + 1}) + \mco(\sqrt{\e}) \ , \\
\cG_\text{bulk}(\De_{\ph^6}; \x) &= 3 \left( \log(\sqrt{\x} + \sqrt{\x + 1}) - \sqrt{\frac{\x}{\x + 1}} \right) + \mco(\sqrt{\e}) \ .
\end{aligned}
\end{equation}
Since $F_{(1)}(\x)$ in \eqref{dyn e} has no $\log(\x)$-term we immediately find
\begin{equation}
\begin{aligned}
\g_{\ph^2} &= \mco(\e^\frac{3}{2}) \ ,
\end{aligned}
\end{equation}
which has to be the case as there are no non-trivial Feynman diagrams for the bulk $\ph^2 - \ph^2$ correlator at $\mco(g^*)$ in the homogeneous theory without a boundary. This means that \eqref{dyn e} is a linear combination of the three blocks above at $\mco(\e^0)$
\begin{equation}
\begin{aligned}
\sqrt{\x}\,F_{(1)}(\x) &= \sum_{n = 1}^{3}\de\la_n\cG_\text{bulk}(n; \x) \ ,
\end{aligned}
\end{equation}
with the bulk OPE coefficients
\begin{equation}
\begin{aligned}
\de\la_1 &= \frac{(N + 4)(N + 2)g^*}{256\,\pi^2\e} - \frac{(N + 2)(\la^*_\pm)^2}{4\,\pi^2\e} \ , \\ 
\de\la_2 &= \frac{(N + 4)(N + 2)g^*}{256\,\pi^2\e} \ , \\
\de\la_3 &= \frac{(N + 4)(N + 2)g^*}{768\,\pi^2\e} \ .
\end{aligned}
\end{equation}
Note that there is no mixing in the bulk-channel due to the e.o.m. \eqref{eom and bc}.

To decompose in boundary blocks we need to look at the full $F(\x)$ at \eqref{e exp dyn}. We can expand the conformal block from the free theory (for $\hp$) in $\e$ using the Mathematica package \textit{HypExp} \cite{Huber:2005yg, Huber:2007dx}. By expanding around large $\x$ we can then read off the same anomalous dimension at \eqref{Bdy anom dim} as well as the following BOE coefficents
\begin{equation} \label{no log BOE}
\begin{aligned}
\de\m_0 &= \frac{(N + 2)(N + 4)g^*}{256\,\pi^2\e}\log 4 - \frac{(N + 2)(\la_\pm^*)^2}{4\,\pi^2\e}  \ , \\
\de\m_1 &= -\frac{(N + 2)(\la_\pm^*)^2}{16\,\pi^2\e} \ .
\end{aligned}
\end{equation}
Due to the b.c. \eqref{eom and bc} there is no mixing between boundary operators. Note that the $\p_\perp\hp$-exchange ($\de\m_1$) arises due to the boundary-interaction.

At the next order there are five Feynman diagrams for the $\ph - \ph$ correlator at $\mco(\e^{\frac{3}{2}})$ (three at $\mco(g\,\la)$ and two at $\mco(\la^3)$). To find this correlator using the e.o.m. we need to calculate two $\ph^5 - \ph$ diagrams at $\mco(\sqrt{\e})$ and four $\hp^3 - \ph$ diagrams at $\mco(\e)$ (two at $\mco(g)$ and two at $\mco(\la^2)$). To avoid the calculation of all of these (three-loop) Feynman diagrams we could instead try to bootstrap the theory using the discontinuity method from \cite{Bissi:2018mcq, Dey:2020jlc, SoderbergRousu:2023nvd}. Let us briefly comment on issues with this approach below.

Since we want to preserve $O(N)$-symmetry, we expect the exchanged bulk primaries to contain an even amount of $\ph$'s (as we have already seen at $\mco(\e)$). Schematically they would be on the form $\mco_n \equiv \p_\m^{2\,n_1}\ph^{2\,n_2}$ with $n_1\in\Z_{\geq 0}$, $n_2\in\Z_{\geq 1}$ and $n \equiv 2\,n_1 + n_2$ (the exact location of the derivatives are not specified). At $\mco(\e^0)$, the corresponding scaling dimensions are integers 
\begin{equation}
\begin{aligned}
\De_n^{(0)} = 2(n_2\,\De_\ph^{(0)} + n_1) = n \in \Z_{\geq 1} \ .
\end{aligned}
\end{equation}
Following \cite{SoderbergRousu:2023nvd}, we study the discontinuity along $\x < -1$ of the bulk blocks hoping to find an orthogonality relation
\begin{equation}
\begin{aligned}
\underset{\x < -1}{\disc}\cG_{\text{bulk}}(n; \x) \propto P_{\frac{n - 3}{2}}^{(\frac{1}{2}, 0)}(t) \ .
\end{aligned}
\end{equation}
However, this Jacobi polynomial is not orthogonal since the argument $\frac{n - 3}{2} \in \frac{\Z_{\geq - 1}}{2}$ is not strictly an integer or a half-integer for $n \in \Z_{\geq 1}$.\footnote{We also to expressed this discontinuity in terms of other polynomials, e.g. Chebyshev or Gegenbauer, but we always found them to be non-orthogonal for the same reason.}

On the other hand, we could try to project out BOE coefficients by studying the discontinuity along $\x \in (-1 ,0)$. As discussed in \cite{SoderbergRousu:2023nvd}, the discontinuity of the boundary block will contain two hypergeometric functions if normal derivatives in $d = 3$ are exchanged (or operators with half-integer scaling dimensions: $\hD^{(0)}_m \in \Z_\geq{0} + \frac{1}{2}$). This makes it difficult to find an orthogonality relation for this discontinuity.\footnote{We encounter the same problem with multiple ${}_2F_1$'s if we assume exchanged boundary operators with scaling dimensions $\hD^{(0)}_m \in \frac{\Z_\geq{1}}{2}$ or $\hD^{(0)}_m \in \Z_\geq{1}$.}

All and all, it seems like the discontinuity method is ill-suited near three dimensions. 
However, there exist other analytical bootstrap methods, such as functional bootstrap \cite{Kaviraj:2018tfd, Mazac:2018biw} and the dispersion relation \cite{Bianchi:2022ppi} that might be more successful in this case.

\subsection{CFT data}

Here we present a summary of the CFT data found upto $\mco(\e)$. Firstly, we found that the bulk fields $\ph$ and $\ph^2$ does not recieve any anomalous dimensions at this order. Furthermore, we normalized the bulk OPE coefficient for the identity exchange to be
\begin{equation}
\begin{aligned}
\la^{\ph\ph}{}_{\1} = 1 \ , \quad \text{(exactly).}
\end{aligned}
\end{equation}
At the bulk tricritical point \eqref{tri fp}, $\hp$ receives an anomalous dimension
\begin{equation}
\begin{aligned}
\De_{\hp} &= \frac{1 - \e}{2} - \frac{(N + 2)(N + 4)}{32(3\,N + 22)}\e + \mco(\e^\frac{3}{2}) \ ,
\end{aligned}
\end{equation}
which is in agreement with \cite{eisenriegler1988surface}. In addition to this, we have the bulk OPE coefficients
\begin{equation}
\begin{aligned}
\la^{\ph\ph}{}_{\ph^{2}}\m^{\ph^{2}}{}_\1 &= 1 + \frac{(N - 24)(N + 2)(N + 4)}{16(N + 8)(3\,N + 22)}\e + \mco(\e^\frac{3}{2}) \ , \\ 
\la^{\ph\ph}{}_{\ph^{4}}\m^{\ph^{4}}{}_\1 &= \frac{(N + 2)(N + 4)}{16(3\,N + 22)}\e + \mco(\e^\frac{3}{2}) \ , \\ 
\la^{\ph\ph}{}_{\ph^{6}}\m^{\ph^{6}}{}_\1 &= \frac{(N + 2)(N + 4)}{48(3\,N + 22)}\e + \mco(\e^\frac{3}{2}) \ ,
\end{aligned}
\end{equation}
and the BOE coefficients
\begin{equation}
\begin{aligned}
(\m^\ph{}_{\hp})^2 &= 2 + \frac{(N + 2)(N + 4)}{3\,N + 22} \left( \frac{\log 2}{8} - \frac{2}{N + 8} \right) \e + \mco(\e^\frac{3}{2}) \ , \\
(\m^\ph{}_{\p_\perp\hp})^2 &= \frac{(N + 2)(N + 4)}{2(N + 8)(3\,N + 22)} \e + \mco(\e^\frac{3}{2}) \ .
\end{aligned}
\end{equation}
The $\ph - \ph$ correlator (on the form \eqref{corr 2}) is given by
\begin{equation*}
\begin{aligned}
F(\x) &= \frac{1}{\x^{\frac{1 - \e}{2}}} + \frac{1}{(\x + 1)^{\frac{1 - \e}{2}}} + \e \left( \frac{(N + 2)(N + 4)}{16(3\,N + 22)} \left( \frac{1}{\sqrt{\x}} + \frac{1}{\sqrt{\x + 1}} \right) \log(\sqrt{\x} + \sqrt{\x + 1}) + \rig\\
\eq\lef - \frac{2(N + 2)(N + 4)}{(N + 8)(3\,N + 22)}\frac{1}{\sqrt{\x + 1}} \right) + \mco(\e^2) \ .
\end{aligned}
\end{equation*}
At the LR f.p. \eqref{LR fp}, when $g^* = 0$, $\hp$ does not receive any anomalous dimension upto $\mco(\e^2)$. This is in agreement with \cite{Prochazka:2019fah}. At this f.p. only $\ph^2$ is exchanged in the bulk-channel, and the non-trivial OPE coefficients are
\begin{equation}
\begin{aligned}
\la^{\ph\ph}{}_{\ph^{2}}\m^{\ph^{2}}{}_\1 &= 1 - \frac{4(N + 2)}{(N + 8)^2}\e^2 + \mco(\e^3) \ , \\ 
(\m^\ph{}_{\hp})^2 &= 2 - \frac{4(N + 2)}{(N + 8)^2}\e^2  + \mco(\e^3) \ , \\
(\m^\ph{}_{\p_\perp\hp})^2 &= \frac{(N + 2)}{(N + 8)^2}\e^2 + \mco(\e^3) \ .
\end{aligned}
\end{equation}
Finally, the $\ph - \ph$ the correlator is
\begin{equation}
\begin{aligned}
F(\x) &= \frac{1}{\x^{\frac{1 - \e}{2}}} + \frac{1}{(\x + 1)^{\frac{1 - \e}{2}}} - \frac{4(N + 2)}{(N + 8)^2}\frac{\e^2}{\sqrt{\x + 1}} + \mco(\e^3) \ .
\end{aligned}
\end{equation}

\section{Coleman-Weinberg mechanism} \label{Sec: CW def}

In this Section we will explain how the CW mechanism works for a general defect. We will then specify to a boundary, where we make contact with \cite{Prochazka:2019fah} which only considers boundary-interaction. After that we apply it to the $\ph^6 - \hp^4$ model, which in addition also has a bulk-interaction.

\subsection{A general defect} \label{Sec: CW gen def}


Let us consider a Euclidean scalar field theory in the presence of a flat $p$-dimensional defect\footnote{The potentials are defined with plus signs since $S$ describes an energy in Euclidean statistical physics.} 
\begin{equation} \label{CWDef action}
\begin{aligned}
S = \int_{\R^d}d^dx \left( \frac{(\p_\m\ph)^2}{2} + V(\ph) \right) + \int_{\R^p}d^px_\para \hat{V}(\hp) \ ,
\end{aligned}
\end{equation}
with a potential in both the bulk, $V(\ph)$, and on the defect, $\hat{V}(\hp)$. Note that $\hat{V}(\hp)$ is not always known, wherein such case we cannot apply the CW mechanism using the technology in this Section. The first thing we want to do is to expand $\ph$ around a classical background (in both the bulk, $\ph_{cl}$, and on the defect, $\hp_{cl}$)
\begin{equation}
\begin{aligned}
\ph = \ph_{cl} + \hslash\,\de\ph \ , \quad \hp = \hp_{cl} + \hslash\,\de\hp \ .
\end{aligned}
\end{equation}
Terms linear in $\de\ph$ and $\de\hp$ in the action gives us the e.o.m. and the b.c. respectively. It brings the action onto the form
\begin{equation} \label{CW def vary action}
\begin{aligned}
S[\ph, \hp] &= S[\ph_{cl}, \hp_{cl}] + \hslash^2\de S[\ph_{cl}, \hp_{cl}, \de\ph, \de\hp] + \mco(\hslash^3) \ , \\
\de S[\ph_{cl}, \hp_{cl}, \de\ph, \de\hp] &= \de S_\text{bulk}[\ph_{cl}, \de\ph] + \de S_\text{def}[\hp_{cl}, \de\hp] \ ,
\end{aligned}
\end{equation}
\begin{equation} \label{CW def vary action 2}
\begin{aligned}
\de S_\text{bulk}[\ph_{cl}, \de\ph] &= \int_{\R^d}d^dx \left( \frac{(\p_\m\de\ph)^2}{2} + \frac{m^2(\ph_{cl})}{2}\de\ph^2 \right) \ , \\
\de S_\text{def}[\hp_{cl}, \de\hp] &= \int_{\R^p}d^px_\para \frac{\hm(\hp_{cl})}{2}\de\hp^2 \ .
\end{aligned}
\end{equation}
We are interested in the one-loop corrections to the effective potentials, thus we only keep terms upto $\mco(\hslash^2)$. For the rest of this Section we use units s.t. $\hslash = 1$. Keeping $\ph_{cl}$ fixed, the action for $\de\ph$ is that of a scalar field theory with different masses in the bulk, $m^2$, and on the defect, $\hm$, (keeping $\ph_{cl}$ and $\hp_{cl}$ fixed)
\begin{equation}
\begin{aligned}
m^2(\ph_{cl}) = \frac{\p^2}{\p\ph_{cl}^2}V(\ph_{cl}) \ , \quad \hm(\hp_{cl}) = \frac{\p^2}{\p\hp_{cl}^2}\hat{V}(\hp_{cl}) \ .
\end{aligned}
\end{equation}
Note that this is a slight abuse of notation as $\hm$ does not necessarily have the correct units of mass (depending on $p$). The quantum fluctuations, $\de\ph$, satisfy the e.o.m. and the b.c.\footnote{In the case of a codimension one defect there might also be an additional contribution to this b.c. coming from the partial integration of the bulk terms. See e.g. \eqref{eom and bc}.}
\begin{equation} \label{delta phi eom}
\begin{aligned}
D^{-1}[\de\ph] &\equiv (-\p_\m^2 + m^2)\de\ph = 0 \ , \quad \text{b.c.}[\de\hp] \equiv \hm\,\de\hp = 0 \ .
\end{aligned}
\end{equation}
Path integration of DQFT's has been worked out in \cite{li1991fluctuation, li1992fluctuation}. In these works they assume a Gaussian theory with several curved defects inside a curved bulk. In path integrating out $\de\ph$ in \eqref{CW def vary action} we are interested in the case of a single flat defect in a flat spacetime. To do this we write the defect contribution as a dirac $\de$-function added to the path integral
\begin{equation} \label{CW def PI}
\begin{aligned}
Z &= \int\D\ph\,e^{-S[\ph]} = \int\D\ph_{cl}\,e^{-S[\ph_{cl}, \hp_{cl}]}\de Z[\ph_{cl}, \hp_{cl}] \ , \\
\de Z &= \int\D\de\ph\,e^{-\de S} = \int\D\de\ph\,\de(\text{b.c.}[\de\hp])e^{-\de S_\text{bulk}} \\
&= \int\D\de\ph\,e^{-\de S_\text{bulk}}\int\D\eta\,\exp\left( -\int_{\R^p}d^px_\para \eta\,\text{b.c.}[\de\hp] \right) \ .
\end{aligned}
\end{equation}
If we complete the square w.r.t. $\de\ph$ we find\footnote{In completing the square we used
	\begin{equation*}
	\begin{aligned}
	\eta\,\text{b.c.}[\hp] &= (-\p_\m^2 + m^2)D_{b.c.}(s_\para, 0, 0)\text{b.c.}[\hp]\,\eta = (-\p_\m^2 + m^2)\text{b.c.}[\hp]\,D_{b.c.}(s_\para, 0, 0)\eta \ ,
	\end{aligned}
	\end{equation*}	
	where we commuted the propagator (a function), $D_{b.c.}(s_\para, 0, 0) \equiv (-\p_\m^2 + m^2)^{-1}$, with $\text{b.c.}[\hp]$.}
\begin{equation}
\begin{aligned}
\de Z &= Z_{\de\ph} Z_\eta \ , \\
Z_{\de\ph} &\equiv \int\D\de\ph\,e^{-\de S_\text{bulk}} \ , \\
Z_\eta &\equiv \int\D\eta\,\exp\left( \int_{\R^p}d^px_\para\int_{\R^p}d^py_\para \eta(x_\para)D_{b.c.}(s_\para, 0, 0)\eta(y_\para) \right) \ ,
\end{aligned}
\end{equation}
where $D_{b.c.}(s_\para, x_\perp, y_\perp)$ is the bulk $\de\ph$-$\de\ph$ correlator subject to the b.c.'s on the defect. Its defect-limit appears in $Z_\eta$. This is a function and is thus not affected by path integrating out $\de\ph$ in $Z_{\de\ph}$. Since we introduced the b.c. on the defect as a Dirac $\de$-function in the path integral \eqref{CW def PI}, we can ignore the effects of the defect in $Z_{\de\ph}$.\footnote{This is in particular important when calculating the trace in the functional determinant.} This means that $Z_{\de\ph}$ is on the same form as without a defect, giving us a similar effective potential in the bulk \cite{PhysRevD.7.1888}
\begin{equation} \label{CW def bulk res}
\begin{aligned}
Z_{\de\ph} &\propto \exp\left(-\int_{\R^d}d^dx\int_{\R^d}\frac{d^dk}{(2\,\pi)^d}\frac{\log (G^{-1})^{ij}(k)}{2} \right) = \exp\left( -\int_{\R^d}d^dx \frac{P_d(m^2, 0)}{4} \right) \ .
\end{aligned}
\end{equation}
Sometimes we keep $\ph_{cl}$ fixed in the CW mechanism, giving us an overall factor of $\text{vol}(R^d)$ which we do not care about. However, this is no longer possible in the presence of a defect, and $\ph_{cl}$ will depend on the normal coordinates, $x_\perp^i$. Above path integral is expressed in terms of the master integral
\begin{equation} \label{Master Int P 0}
\begin{aligned}
P_{n}(m^2, \hm) \equiv \int_{\R^n}\frac{d^nk}{(2\,\pi)^n}\log(\sqrt{k^2 + m^2} + \hm) \ .
\end{aligned}
\end{equation}
This master integral is divergent, but we regularize it using polar coordinates and introducing a large momentum cutoff $\La \gg 1$ for the radius 
\begin{align} \label{Master Int P}
P_{n}(m^2, \hm) &= \frac{S_n}{(2\,\pi)^n} \int_{0}^\La dr\, r^{n - 1}\log(\sqrt{r^2 + m^2} + \hm) \\
&= \frac{\La^{n + 2}}{n(n + 2)(m^2 - \hm^2)} \left[ \frac{\hm}{\sqrt{m^2}} \,F_1\left( \frac{n}{2} + 1; \frac{1}{2}, 1; \frac{n}{2} + 2; -\frac{\La^2}{m^2}, \frac{\La^2}{\hm^2 - m^2} \right) + \rig \nonumber \\
\eq \lef + {}_2F_1\left( 1, \frac{n}{2} + 1; \frac{n}{2} + 2; \frac{\La^2}{\hm^2 - m^2} \right) \right] + \frac{\La^n}{n}\log(\sqrt{\La^2 + m^2} + \hm) \ . \nonumber
\end{align}
Here $F_1(a; b_1, b_2; c; x, y)$ is an Appell $F_1$-series, ${}_2F_1$ is a hypergeometric function, and $S_n$ is the solid angle in $n$ dimensions
\begin{equation}
\begin{aligned}
S_n = \frac{2\,\G_{\frac{n}{2}}}{\pi^{\frac{n}{2}}} \ .
\end{aligned}
\end{equation}
Due to the $F_1$ in \eqref{Master Int P 2}, we cannot directly expand above function around large $\La$ (or for that matter small $m^2$). This forces us to first expand around small $\hm$. After that we are able to expand around large values of $\La$ (or alternatively around small values of $m^2$)
\begin{align} \label{Master Int P 2}
P_{n}(m^2, \hm) &= \frac{S_n}{(2\pi)^n} \left( \La^n \left( \frac{\log\La}{n} - \frac{1}{n^2} + \frac{\hm}{(n - 1)\La} + \frac{m^2 - \hm^2}{2(n - 2)\La^2} + \rig\rig \nonumber \\
\eq  \lef\lef - \frac{m^2\hm}{2(n - 3)\La^3} - \frac{m^2(m^2 - 2\,\hm^2)}{4(n - 4)\La^4} + ... \right) \rig \\
\eq \lef + \frac{\pi}{2}\csc\left(\frac{\pi\, n}{2}\right) \left( \frac{(m^2)^{\frac{n}{2}}}{n} - \frac{(m^2)^{\frac{n}{2} - 1}\hm^2}{2} \right) + \frac{\G_{\frac{1 - n}{2}}\G_{\frac{n}{2}}}{2\sqrt{\pi}}(m^2)^{\frac{n - 2}{2}}\hm \right) \ . \nonumber
\end{align}
Finally we can perform an expansion in $\e$.

Let us now path integrate out $\eta$. Following the steps in \cite{PhysRevD.7.1888}
\begin{equation}
\begin{aligned}
Z_\eta &\propto \sqrt{\det\,D_{b.c.}(s_\para, 0, 0)} = \exp\left( + \frac{\tr\log\,D_{b.c.}(s_\para, 0, 0)}{2} \right) \\
&= \exp\left( \int_{\R^p}\frac{d^ps_\para}{2}\langle s_\para|\log\, D_{b.c.}(s_\para, 0, 0) |s_\para\rangle \right) \ .
\end{aligned}
\end{equation}
If we Fourier transform the states
\begin{equation} \label{CW def FT}
\begin{aligned}
|s_\para\rangle = \int_{\R^p}\frac{d^pk_\para}{(2\,\pi)^p}e^{-i\,k_\para s_\para}|k_\para\rangle \ ,
\end{aligned}
\end{equation}
we find
\begin{equation} \label{CW def def res}
\begin{aligned}
Z_\eta &\propto \exp\left( \int_{\R^p}d^px_\para\int_{\R^p}\frac{d^pk_\para}{(2\,\pi)^p}\frac{\log\, G_{b.c.}(k_\para, 0, 0)}{2} \right) \ .
\end{aligned}
\end{equation}
Here $G_{b.c.}(k_\para, x_\perp, y_\perp)$ is the momentum propagator (w.r.t. the parallel directions) subject to the b.c. of the defect. This is not known for general defects. However, in the specific case of a boundary, this quantity has been worked out in App. A.2 of \cite{Prochazka:2020vog}. We will thus revisit this path integral after having specified to a boundary in Sec. \ref{Sec: CW bdy}. 

Notice the similarity between the bulk contribution \eqref{CW def bulk res} to the effective potential and that on the defect \eqref{CW def def res}. Together they give (upto one-loop)
\begin{equation}\label{CW def V eff}
\begin{aligned}
V_\text{eff}(\ph_{cl}) &= V(\ph_{cl}) + \frac{P_d(m^2, 0)}{4} + \text{c.t.'s} + ... \ , \\
\hat{V}_\text{eff}(\hp_{cl}) &= \hat{V}(\hp_{cl}) - \int_{\R^p}\frac{d^pk_\para}{(2\,\pi)^p}\frac{\log\, G_{b.c.}(k_\para, 0, 0)}{2} + \text{c.t.'s} + ... \ ,
\end{aligned}
\end{equation}
where 'c.t.'s' are the counter-terms (they differ in the bulk and on the defect). The defect potential can be renormalized by introducing an RG scale by defining the defect couplings from derivatives of the classical potentials (we will see an example of this in Sec. \ref{Sec: eom CW}), which in turn might give us a minima of the defect-local fields
\begin{equation} \label{Boundary vev}
\begin{aligned}
\langle\hp_{cl}\rangle = \m^{\De_{\hp}} \ .
\end{aligned}
\end{equation}
We find the bulk v.e.v. using the DOE. Note that we are not at a conformal f.p. in RG. In particular, this changes the operators exchanged in the DOE as well as its differential operator. However, to lowest order in $x_\perp$ we have (assuming the identity is not exchanged)\footnote{Here we rescaled $\hp_{cl}$ s.t. the DOE coefficient for its exchange does not appear.}
\begin{equation} \label{Bulk vev}
\begin{aligned}
\langle\ph_{cl}\rangle = \frac{\langle\hp_{cl}\rangle}{x_\perp^{\De_\ph - \De_{\hp}}} + ... = \m^{\De_{\hp}}x_\perp^{\g_{\hp}} + ... \ ,
\end{aligned}
\end{equation}
which is given in terms of the defect anomalous dimension, $\g_{\hp}$, of $\hp$. This might induce a SSB of the global symmetries in the bulk, even though its corresponding effective potential has not received any radiative one-loop corrections. An example of this phenomena was studied in \cite{Prochazka:2020vog} where no bulk potential was considered. Another example of this, with a bulk potential, will be studied in Sec. \ref{Sec: eom CW}.

\subsection{A boundary} \label{Sec: CW bdy}

To apply the CW mechanism to a boundary, we need the momentum propagator, $G_{b.c.}(k_\para, 0, 0)$, in \eqref{CW def def res}. The corresponding correlator, $D_{b.c.}(s_\para, x_\perp, y_\perp)$ with $s_\para \equiv x_\para - y_\para \in \R^{d - 1}$, in Euclidean space satisfy the KG equation \eqref{delta phi eom} for a massive scalar
\begin{equation}
\begin{aligned}
(-\p_\m^2 + m^2)D_{b.c.}^{ij}(x, y) = \de^{(d)}(s_\para, x_\perp - y_\perp) + \oo\,\de^{(d)}(s_\para, x_\perp + y_\perp) \ ,
\end{aligned}
\end{equation}
satisfying Robin b.c.'s
\begin{equation}
\begin{aligned}
\lim\limits_{x_\perp\rightarrow 0}(\p_{x_\perp} - \hm)D_{b.c.}^{ij}(x, y) = 0 \ .
\end{aligned}
\end{equation}
The correlator satisfying this b.c. was found in App. A.1 of \cite{Prochazka:2020vog} by adding an infinite amount of images on the other side of the boundary. Although not on a closed form, the correlator was found to be given by
\begin{equation}
\begin{aligned}
D_{b.c.}^{ij}(s_\para, x_\perp, y_\perp)  &= D^{ij}(s_\para, x_\perp - y_\perp) + D^{ij}(s_\para, x_\perp + y_\perp) + \\
\eq - 2\,\hm\int_{0}^{\infty}dz\,e^{-\hm\, z}D^{ij}(s_\para, x_\perp + y_\perp + z) \ .
\end{aligned}
\end{equation}
This is on the same form as the wave function in quantum mechanics on a half-line \cite{farhi1990functional}. It reduces down to the correct result for Neumann/Dirichlet b.c. in the limits $\hm \rightarrow 0$ and $\hm \rightarrow \infty$ respectively.\footnote{Note that the results of \cite{Prochazka:2020vog} does not hold for $\hm \rightarrow -\infty$ (see its eq. 60). To study the positive large $\hm$ limit we partially integrate s.t. the overall factor of $\hm$ vanishes. We can then see that only the $z = 0$ limit of the integrand survives
	\begin{equation*}
	\begin{aligned}
	\hm\int_{0}^{\infty}dz\,e^{-\hm\, z}D^{ij}(s_\para, x_\perp + y_\perp + z) &= D^{ij}(s_\para, x_\perp + y_\perp) + \int_{0}^{\infty}dz\,e^{-\hm\, z}\p_zD^{ij}(s_\para, x_\perp + y_\perp + z) \\
	&\stackrel{\hm\rightarrow +\infty}{\longrightarrow} D^{ij}(s_\para, x_\perp + y_\perp) \ .
	\end{aligned}
	\end{equation*}
}

The corresponding momentum propagator (Fourier transformed w.r.t. $s_\para$) was found in Sec. A.2 of \cite{Prochazka:2020vog}. Due to its complicated form, we only write its boundary limit here
\begin{equation} \label{bdy mom prop}
\begin{aligned}
G_{b.c.}^{ij}(k_\para, 0, 0) = \frac{\de^{ij}}{\sqrt{k_\para^2 + m^2} + \hm} \ .
\end{aligned}
\end{equation}
This brings the contribution to the effective potential \eqref{CW def V eff} onto the form
\begin{equation}
\begin{aligned}
V_\text{eff}(\ph_{cl}) &= V(\ph_{cl}) + \frac{P_d(m^2, 0)}{4} + \text{c.t.'s} + ... \ , \\
\hat{V}_\text{eff}(\hp_{cl}) &= \hat{V}(\hp_{cl}) + \frac{P_{d - 1}(m^2, \hm)}{2} + \text{c.t.'s} + ... \ ,
\end{aligned}
\end{equation}
where we remind the reader that $P_{n}(m^2, \hm)$ is the master integral \eqref{Master Int P}. Note that we need to expand it around small values of $\hm$ \eqref{Master Int P 2}. 

The effective potentials above are on the same form as that derived in \cite{Prochazka:2020vog}. In that work, the CW mechanism for a $d = 3 - \e$ dimensional BCFT with only a boundary potential ($m^2 = 0$, $\hm \ne 0$) was developed. In particular, this did not yield an effective potential in the bulk, $V_\eff = 0$, but only one on the boundary. This gives us a non-trivial v.e.v. \eqref{Boundary vev} on the boundary, which extends into the bulk using the BOE \eqref{Bulk vev}. It leads to a SSB of the global symmetry in both the bulk and on the boundary.

In this case, the relevant master integral \eqref{Master Int P} simplifies
\begin{equation}
\begin{aligned}
P_n(0, \hm) &= \frac{S_n\La^n}{(2\,\pi)^nn} \left( \frac{\La^2}{(n + 2)\hm^2} {}_2F_1\left( 1, \frac{n + 2}{2}; \frac{n + 4}{2}; \frac{\La^2}{\hm^2} \right) + \rig \\
\eq \lef - \frac{\La}{(n + 1)\hm} {}_2F_1\left( 1, \frac{n + 1}{2}; \frac{n + 3}{2}; \frac{\La^2}{\hm^2} \right) + \log(\La + \hm) \right) \ .
\end{aligned}
\end{equation}
Its expansion around large $\La$ is
\begin{equation}
\begin{aligned}
P_n(0, \hm) &= \frac{S_n}{(2\,\pi)^n} \left( \La^n \left( \frac{\log\La}{n} - \frac{1}{n^2} + \frac{\hm}{(n - 1)\La} - \frac{\hm^2}{2(n - 2)\La^2} + \rig\rig \\
\eq \lef\lef + \frac{\hm^3}{3(n - 3)\La^3} + ... \right) + \frac{\pi\,\csc(\pi\,n)}{n}\hm^n \right) \ .
\end{aligned}
\end{equation}
In \cite{Prochazka:2020vog} this was further expanded in $\e$ for $n = d - 1$ and $d = 3 - \e$
\begin{equation*}
\begin{aligned}
P_{d - 1}(0, \hm) &= \frac{\hm^2}{4\,\pi} \left( \log\left( \frac{\hm}{\La} \right) - \frac{1}{2} \right) + \frac{\La\,\hm}{8\,\pi} + \frac{\La^2}{4\,\pi} \left( \log\,\La - \frac{1}{2} \right) + ... \ .
\end{aligned}
\end{equation*}

\subsection{Coleman Weinberg mechanism in the $\ph^6 - \hp^4$ model} \label{Sec: eom CW}

In the last part of this paper we will apply the CW mechanism to the $\ph^6 - \hp^4$ model \eqref{phi^6 - phi^4}, and flow along the RG away from the conformal f.p.'s. This is an example when there is both a bulk and a boundary effective potential. We will study the one-loop effects, where only the effective potential on the boundary receives a contribution. 

If we vary the field
\begin{equation}
\begin{aligned}
\ph^i = \ph_{cl}^i + \hslash\,\de\ph^i \ , \quad \hp^i = \hp_{cl}^i + \hslash\,\de\hp^i \ ,
\end{aligned}
\end{equation}
we find (in addition to the e.o.m. and b.c. at \eqref{eom and bc})
\begin{equation} \label{action exp}
\begin{aligned}
S[\ph_{cl}, \de\ph] &= S[\ph_{cl}] + \hslash^2\de S[\ph_{cl}, \de\ph] + \mco(\hslash^3) \ , \\
\de S[\ph_{cl}, \de\ph] &= \int_{\R^d}d^dx \left( \frac{(\p_\m\de\ph^i)^2}{2} + \de V(\ph_{cl}, \de\ph) \right) + \int_{\R^{d - 1}}d^{d - 1}x\, \de \hat{V}(\hp_{cl}, \de\hp) \\
&= \int_{\mathbb{R}^d_+}\frac{d^dx}{2} \de\ph^i (-\p^2|_{b.c.})^{ij} \de\ph^j \ .
\end{aligned}
\end{equation}
From this point on, we set $\hslash = 1$. The potential terms are given by
\begin{equation}
\begin{aligned}
\de V(\ph_{cl}^2, \de\ph^2) = \frac{\de\ph^i(m^2)^{ij}\de\ph^j}{2} \ , \quad  \de \hat{V}(\hp_{cl}, \de\hp) = \frac{\de\hp^i\hm^{ij}\de\hp^j}{2} \ ,
\end{aligned}
\end{equation}
with the bulk and boundary masses for $\de\ph$ (keeping $\ph_{cl}$ constant)
\begin{equation}
\begin{aligned}
(m^2)^{ij} &= m_1^2\de^{ij} + a\,\ph_{cl}^i\ph_{cl}^j \ , \quad &m_1^2 &= \frac{g\,\ph_{cl}^4}{8} \ , \quad &a &= \frac{g\,\ph_{cl}^2}{2} \ , \\
\hm^{ij} &= \hm_1\de^{ij} + \hat{a}\,\hp_{cl}^i\hp_{cl}^j \ , \quad &\hm_1 &= \frac{\la\,\hp_{cl}^2}{2} \ , \quad &\hat{a} &= \la \ .
\end{aligned}
\end{equation}
This brings the differential operator $(-\p^2|_{b.c.})^{ij}$ to the form
\begin{equation}
\begin{aligned}
(-\p^2|_{b.c.})^{ij} &= -\de^{ij}\p^2 + (m^2)^{ij} + \de(x_\perp)\text{b.c.}^{ij} \ , \quad \text{b.c.}^{ij} = -\de^{ij}\p_\perp + \hm^{ij} \ .
\end{aligned}
\end{equation}
Following Sec. \ref{Sec: CW gen def} we write the b.c. as a dirac $\de$-function when we path integrate out $\de\ph$. In the effective potentials for $\ph_{cl}$ we find
\begin{equation} \label{phi6 phi4 CW eff pot}
\begin{aligned}
V_\eff(\ph_{cl}) &\ni -\int_{\R^d}\frac{d^dk}{(2\,\pi)^d}\frac{\tr_{O(N)}\log\,G^{ij}(k)}{2} \ , \\
\hat{V}_\eff(\hp_{cl}) &\ni -\int_{\R^{d - 1}}\frac{d^{d - 1}k_\para}{(2\,\pi)^{d - 1}}\frac{\tr_{O(N)}\log\,G_{b.c.}^{ij}(k_\para, 0, 0)}{2} \ ,
\end{aligned}
\end{equation}
where the trace runs over the $O(N)$-indices. The momentum propagator in the homogeneous theory is given by
\begin{equation} \label{Bulk mom prop}
\begin{aligned}
G^{ij}(k) &= \frac{1}{k^2 + m_1^2} \left( \de^{ij}  - \frac{a\,\ph_{cl}^i\ph_{cl}^j}{k^2 + m_2^2} \right) \ ,
\end{aligned}
\end{equation}
and that satisfying the b.c. is (see App. A.2 in \cite{Prochazka:2020vog} for details on this)\footnote{$H^{ij}$ is found by making the ansatz $H^{ij} = \hat{b}\,\de^{ij} + \hat{c}\,\hp_{cl}^i\hp_{cl}^j$, and then finding the coefficients $\hat{b}$, $\hat{c}$ from $(H^{-1})^{ij}H^{jk} = \de^{ik}$.}
\begin{equation} \label{Bdy mom prop}
\begin{aligned}
G^{ij}_{b.c.}(k_\para, 0, 0) &= H^{ij}_{m_1}(k_\para) + a\,\ph_{cl}^i\ph_{cl}^k\frac{H_{m_1}^{kj}(k_\para) - H_{m_2}^{kj}(k_\para)}{m_1^2 - m_2^2} \ , \\
H^{ij}_{m}(k_\para) &= \left( \hm^{ij} + \sqrt{k_\para^2 + m^2}\de^{ij} \right)^{-1} \\
&= \frac{1}{\sqrt{k_\para^2 + m^2} + \hm_1} \left( \de^{ij} - \frac{\hat{a}\,\hp_{cl}^i\hp_{cl}^j}{\sqrt{k_\para^2 + m^2} + \hm_2} \right)  \ .
\end{aligned}
\end{equation}
Here we defined
\begin{equation}
\begin{aligned}
m_2^2 &\equiv m_1^2 + a\,\ph_{cl}^2 = \frac{5\,g\,\ph_{cl}^4}{8} \ , \quad \hm_2 \equiv \hm_1 + \hat{a}\,\hp_{cl}^2 = \frac{3\,\la\,\hp_{cl}^2}{2} \ .
\end{aligned}
\end{equation}
To find the logarithms in \eqref{phi6 phi4 CW eff pot} we use
\begin{equation}
\begin{aligned}
\log(a\,\1) &= \1\log(a) \ , \\
\log(\1 + b|\ph_{cl}\rangle\langle\ph_{cl}|) 
&= \sum_{n\geq 1}\frac{(-1)^{n + 1}b^n}{n}\ph^{2(n - 1)}|\ph_{cl}\rangle\langle\ph_{cl}| \\
&= \frac{|\ph_{cl}\rangle\langle\ph_{cl}|}{\ph_{cl}^2}\log(1 + b\,\ph_{cl}^2) \ .
\end{aligned}
\end{equation}
The logarithm of the bulk propagator \eqref{Bulk mom prop} is thus
\begin{equation} \label{Log bulk mom prop}
\begin{aligned}
\log G^{ij}(k) &= \log \left( \frac{\de^{jk}}{k^2 + m^2} \right) + \log \left( \de^{jk} - \frac{a\ph^j\ph^k}{k^2 + m_2^2} \right) \\
&= -\de^{jk} \log(k^2 + m_1^2) + \frac{\ph_{cl}^j\ph_{cl}^k}{\ph^2}\log \left( 1 - \frac{a\ph_{cl}^2}{k^2 + m_2^2} \right) \\
&= - \left( \de^{jk} - \frac{\ph_{cl}^j\ph_{cl}^k}{\ph^2} \right) \log(k^2 + m_1^2) - \frac{\ph^j\ph^k}{\ph^2}\log(k^2 + m_2^2) \ ,
\end{aligned}
\end{equation}
giving us the trace in \eqref{phi6 phi4 CW eff pot}
\begin{equation}
\begin{aligned}
&\tr_{O(N)}\log G^{ij}(k) = - (N - 1) \log(k^2 + m_1^2) - \log(k^2 + m_2^2) \ .
\end{aligned}
\end{equation}
Likewise, the trace of the boundary propagator \eqref{Bdy mom prop} is
\begin{equation*}
\begin{aligned}
\tr_{O(N)}\log G_{b.c.}^{ij}(k_\para, 0, 0) &= \log\left(\hm_1^2 + \hm_1 ( \sqrt{k_\para^2 + m_1^2} + \sqrt{k_\para^2 + m_2^2} )  + \sqrt{k_\para^2 + m_1^2}\sqrt{k_\para^2 + m_2^2}\right) + \\
\eq - N\log(\hm_1 + \sqrt{k_\para^2 + m_1^2}) - \log(\hm_1 + \sqrt{k_\para^2 + m_2^2}) + \\
\eq - \log(\hm_2 + \sqrt{k_\para^2 + m_2^2})+ \\
&= -(N - 1)\log(\hm_1 + \sqrt{k_\para^2 + m_1^2}) - \log(\hm_2 + \sqrt{k_\para^2 + m_2^2}) \ .
\end{aligned}
\end{equation*}
All and all, it allows us to express the effective potentials in \eqref{phi6 phi4 CW eff pot} in terms of the master integral (\ref{Master Int P 0}, \ref{Master Int P 2}) (upto one-loop)
\begin{align} \label{eom CW V eff}
V_\text{eff}(\ph_{cl}) &= V(\ph_{cl}) + \frac{(N - 1)P_d(m_1^2, 0) + P_d(m_2^2, 0)}{4} + V_{c.t.}(\ph_{cl}) + ... \ , \\
\hat{V}_\text{eff}(\hp_{cl}) &= \hat{V}(\hp_{cl}) + \frac{(N - 1)P_{d - 1}(m_1^2, \hm_1) + P_{d - 1}(m_2^2, \hm_2)}{2} + \hat{V}_{c.t.}(\hp_{cl}) + ... \ . \nonumber
\end{align}
The $\e$-expansion of the master integrals in $d = 3 - \e$ is
\begin{align} \label{eom CW int}
P_d(m^2, 0) &= -\frac{|m|^3}{12\,\pi} + \frac{m^2\,\La}{4\,\pi^2} + \frac{\La^3}{6\,\pi^2}\left( \log\,\La - \frac{1}{3} \right) + ... \ , \\
P_{d - 1}(m^2, \hm) &= \frac{\hm^2 - m^2}{8\,\pi} \log\left( \frac{m^2}{\La^2} \right) + \frac{m^2}{8\,\pi} - \frac{\hm|m|}{2\,\pi} + \frac{\La^2}{2\,\pi}\left( \log\,\La - \frac{1}{2} \right) + ... \ . \nonumber
\end{align}
The $\La^2$- and $\La^3$-terms are just constants, and thus we will neglect it. The lack of a $\log(m^2)$-term in the bulk is due to the the bulk coupling, $g$, not having a non-trivial RG f.p. at one-loop. In particular this means that the bulk potential will stay the same.

Let us now define the counter-terms. The classical potentials are
\begin{equation}
\begin{aligned}
V(\ph_{cl}) = \frac{g}{48}\ph_{cl}^6 \ , \quad \hat{V}(\hp_{cl}) = \frac{\la}{8}\hp_{cl}^4 \ .
\end{aligned}
\end{equation}
From these we can define the masses (which are zero in our case) and the coupling constants
\begin{equation}
\begin{aligned}
\left.\frac{\p^2V}{\p(\ph_{cl}^2)^2}(\ph_{cl})\right|_{\ph_{cl}^i = 0} &= 0 \ , &\quad \left.\frac{\p^3V}{\p(\ph_{cl}^2)^3}(\ph_{cl})\right|_{\ph_{cl}^i = 0} &= \frac{g}{8} \ , \\
\left.\frac{\p\,\hat{V}}{\p(\hp_{cl}^2)}(\ph_{cl})\right|_{\hp_{cl}^i = 0} &= 0 \ , &\quad \left.\frac{\p^2\hat{V}}{\p(\ph_{cl}^2)^2}(\ph_{cl})\right|_{\hp_{cl}^i = 0} &= \frac{\la}{4} \ .
\end{aligned}
\end{equation}
Based on these (and using the BOE \eqref{Bulk vev}), we introduce the RG scale through
\begin{equation} \label{eom CW def}
\begin{aligned}
\left.\frac{\p^2V_\eff}{\p(\ph_{cl}^2)^2}\right|_{\ph_{cl}^i = 0} &= 0 \ , &\quad \left.\frac{\p^3V_\eff}{\p(\ph_{cl}^2)^3}\right|_{\ph_{cl}^i = \m^{\De_\hp}x_\perp^{\g_{\hp}}\de^{iN} + ...} &= \frac{g}{8} \ , \\
\left.\frac{\p\hat{V}_\eff}{\p(\hp_{cl}^2)}\right|_{\hp_{cl}^i = 0} &= 0 \ , &\quad \left.\frac{\p^2\hat{V}_\eff}{\p(\ph_{cl}^2)^2}\right|_{\hp_{cl}^i = \m^{\De_\hp}\de^{iN}} &= \frac{\la}{4} \ ,
\end{aligned}
\end{equation}
where $\g_{\hp}$ is the boundary anomalous dimension \eqref{Bdy anom dim} (with the coupling constant not tuned to the RG f.p.). 

We define the counter-terms in \eqref{eom CW V eff} as
\begin{equation} \label{eom CW ct}
\begin{aligned}
V_{c.t.} &= A\,\ph_{cl}^4 + \frac{B\,\ph_{cl}^6}{48} \ , \quad \hat{V}_{c.t.} &= \hat{A}\,\ph_{cl}^2 + \frac{\hat{B}\,\hp_{cl}^4}{8} \ .
\end{aligned}
\end{equation}
By implementing \eqref{eom CW def} on the effective potentials \eqref{eom CW V eff} (together with (\ref{eom CW int}, \ref{eom CW ct})) we are able to tune the constants $A$, $B$, $\hat{A}$ and $\hat{B}$ s.t. the divergences in $\La$ vanish. In the bulk we find
\begin{equation}
\begin{aligned}
A = -\frac{(N + 4)g\,\La}{128\,\pi^2} \ , \quad B = \frac{(N + 5\sqrt{5} - 1)|g|^\frac{3}{2}}{16\sqrt{2}\,\pi} \ ,
\end{aligned}
\end{equation}
which completely cancels new addition to the potential in the bulk. On the boundary we find that $\hat{A}$ is given by
\begin{equation}
\begin{aligned}
\hat{A} = -\frac{(N + 2)\la\,\La}{8\,\pi} \ .
\end{aligned}
\end{equation}
The $\hat{B}$ found from \eqref{eom CW ct} is a cumbersome expression. However, it contains
\begin{equation}
\begin{aligned}
\hat{B} \ni \left( (N + 8)\la^2 - \frac{(N + 4)g}{2} \right) \frac{\log\,\La}{2} \ .
\end{aligned}
\end{equation}
This is in fact the divergent part of the bare boundary coupling. By taking a derivative w.r.t. $\log\m$ we find exactly the same $\be$-functions upto one-loop \eqref{eom bdy beta} (under the exchange $\log\,\La \rightarrow \e^{-1}$).\footnote{See e.g. App. B of \cite{Prochazka:2020vog} for details on this.}

\begin{figure} 
	\centering
	\includegraphics[width=0.5\textwidth]{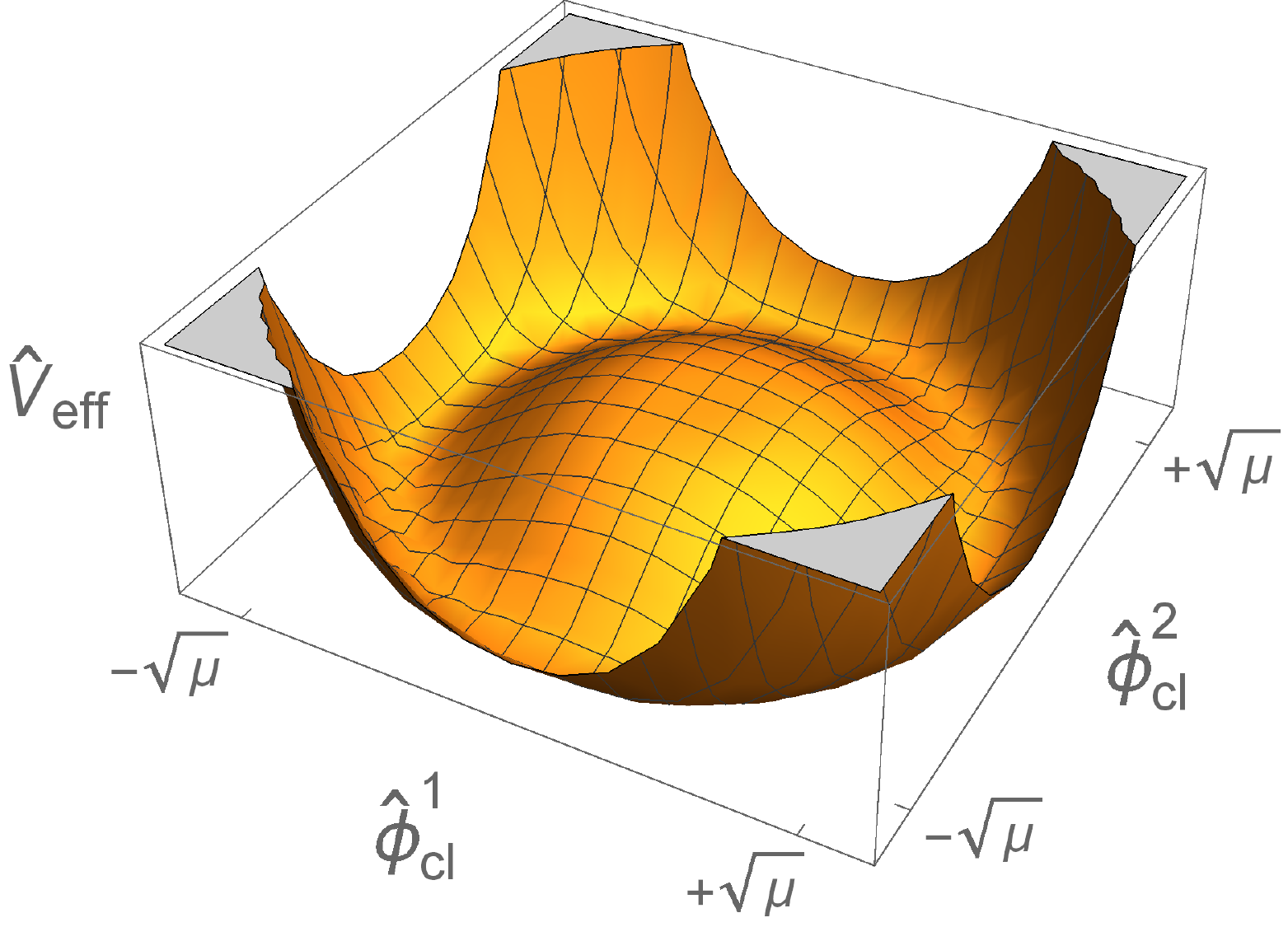}
	\caption{The effective potential, $\hat{V}_\text{eff}(\hp_{cl})$, on the boundary for $N = 2$.}
	\label{Fig: phi6 - phi4 Veff}
\end{figure}

It brings the effective potential to
\begin{equation}
\begin{aligned}
V_\text{eff}(\ph_{cl}) &= V(\ph_{cl}) + ... \ , \\
\hat{V}_\text{eff}(\hp_{cl}) &= \frac{\la\,\hp_{cl}^4}{8} + \left( (N + 8)\la^2 - \frac{(N + 4)g}{2} \right) \frac{\hp_{cl}^4}{2} \left( \log\left( \frac{\hp_{cl}^2}{\m} \right) - \frac{3}{2} \right) + ... \ .
\end{aligned}
\end{equation}
The way we introduced the RG scale, $\m$, in \eqref{eom CW def} tells us that $\hat{V}_\text{eff}$ has a minimum at this point. This gives us a relation between the coupling constants
\begin{equation} \label{RG point}
\begin{aligned}
\left.\hp_{cl}^i\frac{\p\,\hat{V}_\text{eff}}{\p\hp_{cl}^i}\right|_{\hp_{cl}^2 = \m + \mco(\e)} &= 0 \qRq g = -\frac{8\,\pi\,\la}{N + 4} + \frac{2(N + 8)\la^2}{N + 4} \ ,
\end{aligned}
\end{equation}
which does not flow to any of the f.p.'s in the RG flow in Fig. \ref{Fig: RGphi6phi4}. Inserting this into $\hat{V}_\eff$ 
\begin{equation}
\begin{aligned}
\hat{V}_\eff = \frac{\la\,\hp_{cl}^4}{8} \left( \log\left( \frac{\hp_{cl}^2}{\m} \right) - \frac{1}{2} \right) + ...\ .
\end{aligned}
\end{equation}
A plot of the boundary effective potential for $N = 2$ is in Fig. \ref{Fig: phi6 - phi4 Veff}, where we can see that it has an $O(N)$-invariant minima at $\sqrt{\m}$ as in \eqref{Boundary vev}. This means that $\hp_{cl}$ has received a non-trivial v.e.v.
\begin{equation}
\begin{aligned}
\langle\hp_{cl}^i\rangle = \m^{\De_{\hp}}\de^{iN} = \sqrt{\m}\,\de^{iN} + \mco(\e, \la) \ .
\end{aligned}
\end{equation}
As was explained in Sec. \ref{Sec: CW gen def} the BOE will in turn induce a v.e.v. in the bulk \eqref{Bulk vev} (to lowest order in $x_\perp$)
\begin{equation}
\begin{aligned}
\langle\ph_{cl}^i\rangle = \m^{\De_{\hp}}x_\perp^{\g_{\hp}} + ... = \sqrt{\m}\,\de^{iN} + \mco(\e, \la, x_\perp) \ .
\end{aligned}
\end{equation}
This means that a SSB of the $O(N)$-symmetry occurs, leaving us with $O(N - 1)$-symmetry. We can thus apply the Higgs mechanism and expand around this minimum\footnote{This is a slight abuse of notation, as the Higgs mechanism actually explains how the vector bosons in particle physics become massive. A similar phenomena can also be seen in \cite{Prochazka:2020vog}, where a SSB of the global symmetry of $\ph$ generates a mass for the $\ch$-fields (see its eq. (42, 43)).}
\begin{equation} \label{Exp vev}
\begin{aligned}
\ph_{cl}^i &= e^{\eta^kT^k} (\sqrt{\m} + \s)\de^{iN} + \mco(\e, \la, x_\perp) \ , \\
\hp_{cl}^i &= e^{\hat{\eta}^kT^k} (\sqrt{\m} + \hat{\s})\de^{iN} + \mco(\e, \la) \ .
\end{aligned}
\end{equation}
Here $\s$ is the Higgs mode, $\eta^k$ is the massless Goldstone mode and $T^k$ is a generator of the Lie algebra corresponding to the broken part, $O(N)/O(N - 1)$, of the original symmetry group. The number of generators are
\begin{equation}
\begin{aligned}
\text{dim } O(N)/O(N - 1) &= \text{dim } O(N) - \text{dim } O(N - 1) \\
&= \frac{N(N - 1)}{2} - \frac{(N - 1)(N - 2)}{2} = N - 1 \ .
\end{aligned}
\end{equation}
This means that $k \in \{1, ..., N - 1\}$. 

Expanding around the v.e.v. \eqref{Exp vev} yields kinetic terms to $\eta^k$ and $\s$ in the bulk
\begin{equation*} 
\begin{aligned}
S = \int_{\R^d}d^dx \left( \frac{(\p_\m\s)^2}{2} + \frac{(\sqrt{\m} + \s)^2(\p_\m\eta^k)^2}{2} + V_\eff \right) + \int_{\R^{d - 1}}d^{d - 1}x_\para \hat{V}_\eff \ .
\end{aligned}
\end{equation*}
In the low-energy limit ($k_\m \rightarrow 0$) of $\eta^k$, its mixed interactions with $\s$ vanish and thus becomes free. If we expand the effective potentials in $\s$ we find interactions which break the $O(N)$-symmetry down to $O(N - 1)$ both in the bulk and on the boundary
\begin{equation} \label{bulk eff pot}
\begin{aligned}
V_\eff &= \frac{g}{48}\sum_{n = 1}^6 \binom{6}{n} \m^{3 - \frac{n}{2}}\s^n \ , \\
\hat{V}_\eff &= \frac{\la\,\m}{2}\s^2 + \frac{5\,\la\,\sqrt{\m}}{6}\s^3 + \frac{11\,\la}{24}\s^4 - 6\,\la\sum_{n\geq 5}\frac{(-1)^n(n + 1)_{-5}}{\m^{\frac{n}{2} - 2}}\s^n \ ,
\end{aligned}
\end{equation}
where we neglected constant terms. This is an effective field theory for a first-order p.t. The Higgs mode has received the bulk and boundary masses (using the relation \eqref{RG point} between the coupling constants)
\begin{equation}
\begin{aligned}
m_\s^2 = \frac{5\,g\,\m^2}{8} = -\frac{5\,\pi\,\la\,\m^2}{N + 4} \left( 1 - \frac{N + 8}{4\,\pi}\la \right) \ , \quad \hat{m}_{\s} = \la\,\m \ .
\end{aligned}
\end{equation}
For the fields to be physical (with positive energy) we require $m_\s^2 > 0$. This leads to $\la < 0$ (and $g > 0$ due to \eqref{RG point}) if we consider infinitesimal couplings, $\la \ll 1$, and finite values of $N \in \Z_{\geq 1}$. 

\section{Conclusion}

In this paper we found the bulk correlator in the $\ph^6 - \hp^4$ model \eqref{phi^6 - phi^4} from the e.o.m. and b.c. \eqref{eom and bc}. With this correlator at hand we commented on problems with the discontinuity method in odd dimensions. Finally we applied the CW mechanism to this model, giving us an effective description of a first order p.t. \eqref{bulk eff pot}. Here we saw how a v.e.v. on the boundary extends into the bulk, and thus breaks the global $O(N)$-symmetry.

It would be interesting to apply the CW mechanism to a model where both the bulk and the boundary receives a non-trivial contribution to the effective potential. This would require a theory with at least two bulk couplings s.t. we can stay in a perturbative regime (see eq. 3.13 in \cite{PhysRevD.7.1888}). A difficulty with this task is to find a suitable model where the calculations are manageable. Going to higher loops \cite{Tan:1997ew} (in e.g. the $\ph^6 - \hp^4$ model \eqref{phi^6 - phi^4}) is also rather difficult, as we comment on in App. \ref{App: two-loop}.


\section*{Acknowledgement}

I would like to express my gratitude to Agnese Bissi, Miztani Euich, Mykola Shpot and Vladimir Proch\'azka for enriching discussions on the Coleman-Weinberg mechanism. 
I also thank everyone that went to my public defence of my thesis \cite{SoderbergRousu:2023ucv}, where the results in this paper was first presented. This project was funded by Knut and Alice Wallenberg Foundation grant KAW 2021.0170, VR grant 2018-04438 and Olle Engkvists Stiftelse grant 2180108.


\appendix

\section{Coleman-Weinberg mechanism at two-loop} \label{App: two-loop}

\begin{figure} 
	\centering
	\includegraphics[width=0.5\textwidth]{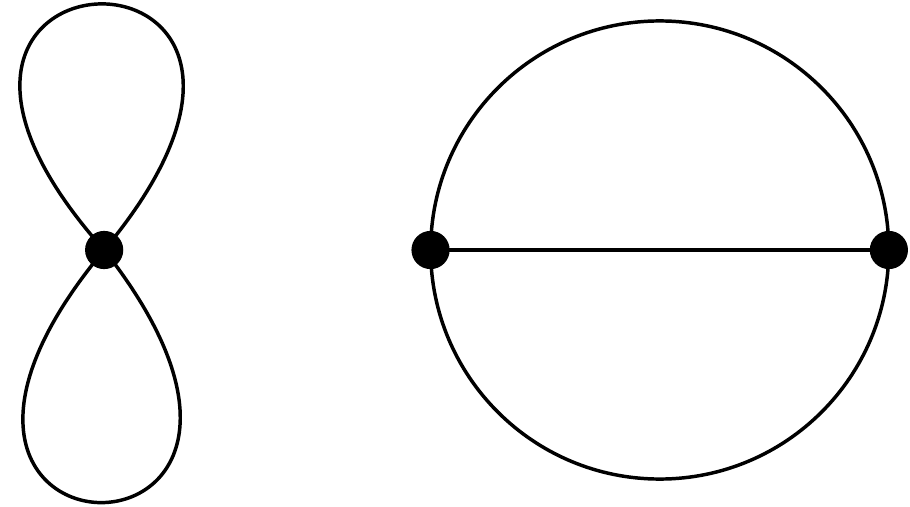}
	\caption{The two diagrams that will contribute to the effective potentials at two-loop.}
	\label{Fig: CW 2-loop}
\end{figure}

In this Appendix we will comment on issues of the CW mechanism at two-loops. To find the two-loop effective potentials we need to consider higher order terms in $\de\ph$ and $\de\hp$ (in e.g. \eqref{action exp}). This leads to a non-Gaussian theory for $\de\ph$, and thus we have to calculate Feynamn diagrams with no external operators and internal $\de\ph - \de\ph$ correlators when we path integrate out $\de\ph$ \cite{Tan:1997ew}. In the bulk contribution to the effective potential, $V_\eff$, the $\de\ph - \de\ph$ correlator is given by \eqref{Bulk mom prop}, and on the boundary, $\hat{V}_\eff$, by \eqref{Bdy mom prop}. Let us here focus on $\hat{V}_\eff$ which is more difficult to find. Already for $\Z_2$-symmetry,\footnote{Or the \textit{edge ordering} when $\ph_{cl}^i = \ph_{cl}\de^{iN}$.} the boundary correlator \eqref{Bdy mom prop} is rather complicated
\begin{equation}
\begin{aligned}
G_{b.c.}(k_\para, 0, 0) &= \frac{1}{\sqrt{k_\para^2 + m^2} + \hm} \frac{\sqrt{k_\para^2 + m^2} - \hm_1}{\sqrt{k_\para^2 + m^2} - \hm} \\
&= \frac{\sqrt{k_\para^2 + m^2}}{k_\para^2 + m^2 - \hm^2} \frac{\sqrt{k_\para^2 + m^2}}{\sqrt{k_\para^2 + m^2}} - \frac{\hm^2}{k_\para^2 + m^2 - \hm^2} \\
&= \frac{k_\para^2 + m^2 - \hm^2 + \hm^2}{\sqrt{k_\para^2 + m^2}(k_\para^2 + m^2 - \hm^2)} - \frac{\hm^2}{k_\para^2 + m^2 - \hm^2} \\
&= P(k_\para) - \hm\,Q(k_\para) + \hm^2 P(k_\para)Q(k_\para) \ ,
\end{aligned}
\end{equation}
which is expressed in terms of
\begin{equation}
\begin{aligned}
P(k_\para) &= \frac{1}{\sqrt{k_\para^2 + m^2}} \ , \quad Q(k_\para) = \frac{1}{k_\para^2 + m^2 - \hm^2} \ .
\end{aligned}
\end{equation}
Note that the $\de\ph^5$- and $\de\ph^6$-terms will not contribute to any non-trivial Feynman diagrams at two-loop, and that it is enough to consider the $\de\ph^3$- and $\de\ph^4$-terms. From these two vertices we find the two non-trivial Feynman diagrams in Fig. \ref{Fig: CW 2-loop}. Let us here focus on the \textit{sunset diagram} (the second one) which is more difficult to calculate. For the boundary contribution with $\Z_2$-symmetry, it will contain the following integral\footnote{For $O(N)$-symmetry, or the \textit{axial ordering}: $\ph_{cl}^i$, $i\in\{1, ..., N\}$, the number of denominator are twelve.}
\begin{equation}
\begin{aligned}
I &\propto \int_{\R^{d - 1}}\frac{d^{d - 1}k_\para}{(2\,\pi)^{d - 1}} \int_{\R^{d - 1}}\frac{d^{d - 1}l_\para}{(2\,\pi)^{d - 1}} G_{b.c.}(k_\para, 0, 0) G_{b.c.}(k_\para + l_\para, 0, 0) G_{b.c.}(l_\para, 0, 0) \\
&\ni \hm^6\int_{\R^{d - 1}}\frac{d^{d - 1}k_\para}{(2\,\pi)^{d - 1}} \int_{\R^{d - 1}}\frac{d^{d - 1}l_\para}{(2\,\pi)^{d - 1}} P(k_\para)Q(k_\para)P(k_\para + l_\para)Q(k_\para + l_\para)P(l_\para)Q(l_\para) \ .
\end{aligned}
\end{equation}
So if we want to study two-loop contributions to the effective potential we need to calculate above integral.

Let us also mention that to avoid the issue of having one bulk coupling we can consider the $H(N)$-model in $d = 3 - \e$ with three bulk couplings \cite{Osborn:2017ucf, BenAliZinati:2021rqc} ($[(\ph^i)^2]^3$, $(\ph^i)^2(\ph^j)^4$ and $(\ph^i)^6$ with implicit summation over the indices $i$, $j \in \{1, ..., N\}$) and two boundary couplings ($[(\ph^i)^2]^2$ and $(\ph^i)^4$). This model is invariant under the hypercubic group $H(N) = S_N \rtimes \Z_2^N$, which is a subgroup of $O(N)$.

\bibliographystyle{utphys}
\footnotesize
\bibliography{References}	
	
\end{document}